\documentstyle[aps,eqsecnum,multicol,epsf]{revtex} 
\begin{document}
\def\Or {\overline{r}}
\def\Obr{\overline{\B.r}}
\def\Btensor{ B_{n,\ell,\sigma}^{\alpha_1 \ldots \alpha_{n}} } 
\def\Ctensor #1#2{ \delta^{\alpha_{#1}\alpha_{#2}} 
B_{n-2,\ell,\sigma}^{
\stackrel{ \text{no } #1 ,#2}
{ \overbrace{_{\alpha_1 \ldots \alpha_{n}} } }}}
\def\I.#1{\it #1}
\def\B.#1{{\bbox#1}}
\def\C.#1{{\cal  #1}}
\def\BE{\begin{equation}}
\def\EE{\end{equation}}
\def\BEA{\begin{eqnarray}}
\def\EEA{\end{eqnarray}}
\def\nn{\nonumber}
\def\hf{\case{1}{2}}
\def\e{\epsilon} 
\def\a{\alpha}
\def\b{\beta}
\def\d{\delta}
\def\g{\gamma}
\def\k{\kappa}
\def\L{\Lambda}
\def\kt{\kappa_{_{\rm T}}}
\def\k0{\kappa_0}
\def\Od{\Omega_d}
\def\la{\left\langle}
\def\ra{\right \rangle}
\def \LA { \Bigg\lfloor} 
\def \RA { \Bigg \rfloor_\B.e}
\def \lv  {{  \bf \Large \{}}
\def \rv  {{  \bf \Large \}}}
\title{
 Anomalous Scaling in the Anisotropic Sectors \\
of the Kriachnan Model  of Passive Scalar Advection}
\author {Itai Arad$^*$, 
Victor L'vov$^{*\dag}$,  Evgenii 
Podivilov$^{*\dag}$ and  Itamar Procaccia$^*$}
\address{$^*$Departments of~~Chemical Physics 
The Weizmann Institute of Science, Rehovot 
76100, Israel, and \\
$^{\dag}$Institute of Automation and Electrometry,
 Ac. Sci. of Russia, 630090, Novosibirsk, Russia}
\maketitle
\begin{abstract}
Kraichnan's model of passive scalar advection in which the driving
(Gaussian) velocity field has fast temporal decorrelation is studied
as a case model for understanding the anomalous scaling behavior in
the anisotropic sectors of turbulent fields. We show here that the
solutions of the Kraichnan equation for the $n$ order correlation
functions foliate into sectors that are classified by the irreducible
representations of the SO($d$) symmetry group. We find a discrete
spectrum of universal anomalous exponents, with a different exponent
characterizing the scaling behavior in every sector.  Generically the
correlation functions and structure functions appear as sums over all
these contributions, with non-universal amplitudes which are determined
by the anisotropic boundary conditions. The isotropic sector is always
characterized by the smallest exponent, and therefore for sufficiently
small scales local isotropy is always restored.  The calculation of
the anomalous exponents is done in two complementary ways. In the
first they are obtained from the analysis of the correlation functions
of {\em gradient fields}. The theory of these functions involves the
control of logarithmic divergences which translate into anomalous
scaling with the ratio of the inner {\em and} the outer scales
appearing in the final result. In the second way we compute the
exponents from the zero modes of the Kraichnan equation for the
correlation functions of the scalar field itself. In this case the
renormalization scale is the outer scale. The two approaches lead to
the same scaling exponents for the same statistical objects,
illuminating the relative role of the outer and inner scales as
renormalization scales. In addition  we derive exact fusion
rules which govern the small scale asymptotics of the correlation
functions in all the sectors of the symmetry group and in all
dimensions.
\end{abstract} 
\pacs{PACS numbers 47.27.Gs, 47.27.Jv, 05.40.+j}
\begin{multicols}{2}
\section{Introduction} \label{intro} 
The aim of this paper is twofold. On the one hand we are interested in
the effects of anisotropy on the universal aspects of scaling behavior
in turbulent systems. To this aim we present below a theory of the
anomalous scaling of the Kraichnan model of turbulent
advection\cite{68Kra} in anisotropic sectors which are classified by
the irreducible representations of the SO($d$) symmetry group. On the
other hand we are interested in clarifying the relationship between
ultraviolet and infrared anomalies in turbulent systems. Again, it
turns out that the Kraichnan model is an excellent case model in which
this relationship can be exposed with complete clarity. As is well
known by now, the Kraichnan model describes the advection of a
passive scalar by a velocity field that is random, Gaussian and
delta-correlated in time. The correlation functions of the field scale
in their spatial dependence, and the main question is what are the
scaling (or homogeneity) exponents of the statistical objects of the
scalar field that are induced by the given value of the scaling
exponent of the advecting velocity field.

The two issues discussed in this paper have an importance that
transcends the particular example that we treat in detail in this
paper. The first is the role of anisotropy in the observed scaling
properties in turbulence. We have shown recently that in the presence
of anisotropic effects (which are ubiquitous in realistic turbulent
systems) one needs to carefully disentangle the various universal
scaling contributions. Even at the largest available Reynolds numbers
the observed scaling behavior is not simple, being composed of several
contributions with different scaling exponents.  The statistical
objects like structure functions and correlations functions are
characterized by one scaling (or homogeneity) exponent only in the
idealized case of full isotropy, or infinite Reynolds numbers when the
scaling regime is of infinite extent. Anisotropy results in mixing in
various contributions to the statistical objects, each of which is
characterized by one universal exponent, but the total is a sum of
such contributions which appears not to ``scale" in standard log-log
plots. By realizing that the correlation functions have natural
projections on the irreducible representations of the SO(3) symmetry
group we could offer methods of data analysis that allow one to
measure the universal scaling exponents in each sector separately. In
this paper we show that this foliation is an exact property of the
statistical objects that arise in the context of the Kraichnan model.

The second issue that transcends the particular example of the
Kraichnan model is the identification of the renormalization scales
that are associated with anomalous exponents. As has been already
explained before, the renormalization scale which appears in the
correlation functions of the passively advected scalar field is the
outer scale of turbulence $L$. On the other hand, the theory of
correlations of {\em gradients} of the field expose the inner (or
dissipative) scale a an additional renormalization scale.  Having
below a theory of anomalous scaling in  various sectors of the
symmetry group allows us to explain clearly the relationship between
the two renormalization scales and the anomalous exponents that are
implied by their existence. Since we expect that Kolmogorov type
theories, which assume that no renormalization scale appears in the
theory, are generally invalidated by the appearance of both the outer
and the inner scales as renormalization scales, the clarification of
the relation between the two is important also for other cases of
turbulent statistics.

The central quantitative result of this paper is the expression for
the scaling exponent $\zeta_{n}^{(\ell)}$ which is associated with the
scaling behavior of the $n$-order correlation function (or structure
function) of the scalar field in the $\ell$'th sector of the symmetry
group. In other words, this is the scaling exponent of the projection
of the correlation function on the $\ell$'th irreducible
representation of the SO($d$) symmetry group, with $n$ and $\ell$
taking on even values only, $n=0,2, \dots$ and $\ell=0,2,\dots$:
\begin{equation}
\zeta_{n}^{(\ell)}=
n-\epsilon\Big[\frac{n(n+d)}{2(d+2)}
-\frac{(d+1)\ell(\ell+d-2)}{2(d+2)(d-1)}\Big]
+O(\epsilon^2) \ .
\end{equation} 
The result is valid for any even $\ell\le n$, and to $O(\epsilon)$
where $\epsilon$ is the scaling exponent of the eddy diffusivity in
the Kraichnan model (and see below for details). In the isotropic
sector ($\ell=0$) we recover the well known result of \cite{96BGK}. It
is noteworthy that for higher values of $\ell$ the discrete spectrum
is a strictly increasing function of $\ell$. This is important, since
it shows that for diminishing scales the higher order scaling
exponents become irrelevant, and for sufficiently small scales only
the isotropic contribution survives. As the scaling exponent appear in
power laws of the type $(R/\Lambda)^\zeta$, with $L$ being some
typical outer scale and $R\ll \Lambda$, the larger is the exponent,
the faster is the decay of the contribution as the scale $R$
diminishes. This is precisely how the isotropization of the small
scales takes place, and the higher order exponents describe the rate
of isotropization. Nevertheless for intermediate scales or for finite
values of the Reynolds and Peclet numbers the lower lying scaling
exponents will appear in measured quantities, and understanding their
role and disentangling the various contributions cannot be avoided.

The organization of this paper is as follows: In Sect.~2 we recall the
Kraichnan model of passive scalar advection, and introduce the
statistical objects of interests.  In Sect.~3 we set up the calculation
of the correlation functions of gradients of the field. It turns out
that it is most straightforward to compute the fully fused correlation
functions of gradient field, as these objects depend only on the ratio
of the outer and inner scales.  We compute these quantities and their
exponents to first order in $\epsilon$. We introduce the appropriate
irreducible representations of the SO($d$) symmetry group and evaluate
the scaling exponents in all its sectors. In Sect. 4 we turn to the
correlations functions of the passive scalar field itself, and compute
the scaling exponents of the structure functions in the presence of
anisotropy, again correct to first order in $\epsilon$. To this aim we
compute the zero modes in all the sectors of the symmetry group. One
of the interesting points of this paper is that the results of this
calculation and the calculation via the correlations of the gradient
fields gives the same results for the scaling exponents if one accepts
the fusion rules. To clarify the issue we prove the fusion rules here
in all the sectors of the symmetry group by a direct computation of
the fusion of the zero modes. In Sect. 5 we offer a summary and a
discussion.
\section{Kraichnan's Model of Turbulent Advection and the Statistical Objects}
\label{s:Kra}
The model of passive scalar advection with rapidly decorrelating
velocity field was introduced by R.H. Kraichnan \cite{68Kra} already
in 1968. In recent years 
\cite{96BGK,94Kra,94LPF,95GK,95CFKL,96FGLP}
it was shown to be a fruitful case model for understanding
multi-scaling in the statistical description of turbulent fields. The
basic dynamical equation in this model is for a scalar field
$T({\B.r},t)$ advected by a random velocity field ${\B.u}({\B.r},t)$:
\begin{equation}
\label{advect}
        \big[\partial_t  - \kappa_0 \nabla^2  + 
        {\B.u}({\B.r},t) \cdot \bbox{\nabla}\big]
         T({\B.r},t) = f({\B.r},t)\ .
\end{equation}
In this equation $f({\B.r},t)$ is the forcing.  In Kraichnan's model
the advecting field ${\B.u}({\B.r},t)$ as well as the forcing field
$f({\B.r},t)$ are taken to be Gaussian, time and space homogeneous,
and delta-correlated in time:
\begin{eqnarray}\label{ff}
        \overline{  f({\B.r},t)  f({\B.r}',t') } &=& 
           \Phi({\B.r}-{\B.r}')  \delta(t - t')\,,\\   
\label{W}
\langle u^\alpha({\B.r},t) u^\beta({\B.r}',t') \rangle
&=&\C.W^{\alpha\beta}({\B.r}-{\B.r}')  \delta(t - t')\,,
\EEA
Here the symbols ~$\overline{\cdots}$~ and ~$\langle \cdots \rangle$~
stand for independent ensemble averages with respect to the statistics
of $f$ and ${\B.u}$ which are given {\em a priori}.  We will study
this model in the limit of large Peclet (Pe) number, Pe$\equiv
U_{\Lambda} \Lambda/\kappa_0$, where $U_\Lambda$ is the typical size
of the velocity fluctuations on the outer scale $\Lambda$ of the
velocity field. We stress that the forcing is {\em not} assumed
isotropic, and actually the main goal of this paper is to study the
statistic of the scalar field under anisotropic forcing.

The correlation function of the advecting velocity needs further
discussion.  It is customary to introduce $\C.W^{\alpha
\beta}(\B.R)$ via its $\B.k$-representation:
\BEA\label{W1}
\C.W^{\alpha\beta}(\B.R)&=&\frac{\e\, D}{\Od}
\int \limits _{\L ^{-1}}^{\lambda^{-1}}     \frac{  d^d p}
{p^{d+\e}} \,
P^{\alpha\beta}(\B.p)\,
\exp (-i \B.p\cdot \B.R)\,,\\
\label{tp}
P^{\alpha\beta}(\B.p) &=&
\Big[\delta_{\alpha\beta}
-\frac {p^\alpha p^\beta}{p^2} \Big] \,,
\EEA
were $P^{\alpha\beta}(\B.p)$ is the transversal projector, $\Od=(d-1)
\Omega(d)/d $ and $\Omega(d)$ is the volume of the sphere in $d$
dimensions ({\em i.e.}  $\Omega(2)=2\pi$, $\Omega(3)=4\pi$). Equation
(\ref{W1}) introduces the four important parameters that determine the
statistics of the driving velocity field: $\L$ and $\lambda$ are the
outer and inner scales of the driving velocity field respectively.
The scaling exponent $\e$ characterizes the correlation functions of
the velocity field, lying in the interval $(0,2)$. The factor $D$ is
related to the correlation function~(\ref{W}) as follows:
\BE\label{W3}
\C.W^{\alpha\beta}(0)=D\delta_{\a\b}(\L^\e -\lambda^\epsilon)\ .
\EE
The most important property of the driving velocity field from the
point of view of the scaling properties of the passive scalar is the
tensor of ``eddy diffusivity" \cite{68Kra}
\BE\label{eddy-diff}
 \kappa^{\alpha\beta}_{_{\rm T}}({\B.R}) 
\equiv   2[\C.W^{\alpha\beta}(0)-\C.W^{\alpha\beta}(\B.R)]\ .
\EE
The scaling properties of the scalar depend sensitively on the scaling
exponent $\e$ that characterizes the $R$ dependence of
$\kappa^{\alpha\beta}_{_{\rm T}}({\B.R})$:
\BEA\label{kappa1} 
\kappa^{\alpha\beta}_{_{\rm T}}({\B.R})&\propto & [\L^\e
 -\lambda^\epsilon]\delta_{\alpha\beta}\,, \quad \mbox {for} \quad
 R\gg \L\,,\\
\kappa^{\alpha\beta}_{_{\rm T}}({\B.R})&\propto & R^\e
 \Big[\delta_{\alpha\beta}
-\frac{\epsilon}{d-1+\epsilon}{R^\alpha
 R^\beta
\over R^2}\Big],        
\quad \lambda\ll R\ll\L \ .\nn
\EEA

We are interested in the scaling properties of the scalar field. By
this we mean the power laws characterizing the $R$ dependence of the
various correlation and response functions of $T({\B.r},t)$ and its
gradients.  We will focus on three types of quantities:

1)    ``Unfused" structure functions  are defined as
\begin{eqnarray}
\label{Sunfused}
S_{n}(\B.r_1,\overline{\B.r}_1,\dots \B.r_{n},\overline{\B.r}_{n})
&\equiv &\langle [T(\B.r_1,t)-T(\overline{\B.r}_1,t)]\\ \nn
\times
[T(\B.r_2,t)-T(\overline{\B.r}_2,t)]&\dots&
[T(\B.r_n,t)-T(\overline{\B.r}_n,t)]\rangle \ ,
\end{eqnarray}
and in particular the standard  structure functions are
\begin{equation}  
\label{sn-def}
        S_{n}({\B.R})   \equiv 
        \langle [ T({\B.r}+{\B.R}, t) - 
        T({\B.r}, t)]^{n} \rangle \ .              
\end{equation}  
In writing this definition  we used the fact that the stationary and
space-homogeneous statistics of the velocity and the forcing fields
lead to a stationary and space homogeneous ensemble of the scalar
$T$. If the statistics is also isotropic, then $S_{n}$ becomes a
function of $R$ only, independent of the direction of ${\B.R}$.  The
``isotropic scaling exponents'' $\zeta_{n}$ of the structure
functions
\begin{equation}
\label{zeta-def}
        S_{n}(R) \propto R^{\zeta_{n}} ,                
\end{equation} 
characterize their $R$ dependence in the limit of large Pe, when $R$
is in the ``inertial" interval of scales. This range is
$\lambda,\eta\ll R\ll\Lambda,L$ where $\eta$ is the dissipative
scale of the scalar field,
\begin{equation}
\eta = \Lambda \left(\frac{\kappa_0}{D}\right)
^{1/\epsilon} \ . \label{eta}
\end{equation}

2) In addition to structure functions we are also interested in the
simultaneous $n$th order correlation functions of the temperature
field which is time independent in stationary statistics:
\BE \label{Fn}
\C.F_{n}(\{\B.r_m \})
\equiv \la T(\B.r_1,t)\,T(\B.r_2,t)\dots
T(\B.r_{n},t)\ra \,,
\EE
were we used the shorthand notation $\{\B.r_m \}$ for the whole set of
arguments of $n$th order correlation function $\C.F_{n}$,
$\B.r_1,\B.r_2\dots \B.r_{n}$.

3)  Finally, we are interested in correlation functions of the
gradient field $\B.\nabla T$. There can be a number of these, and we
denote
\BE\label{Psi1}
\C.H_{n}^{\{\a_m\}}    (\{\B.r_m\})       \equiv
\Big\langle \prod_{i=1}^{n}
\Big[\nabla^{\a_i}T(\B.r_i,t)\Big]\Big\rangle
\,,
\EE
where $\{\a_m\}$ is a set of even $n$ vector indices
$\{\a_m\}=\a_1,\a_2\dots \a_{n}$.  We introduce also one-point
correlations which in the space homogeneous case is independent of
the space coordinates:
\BE\label{Psi2} H_{n}^{\{\a_m\}}
\equiv 
\C.H_{n}^{\a_1\a_2\dots \a_{n}}(\{\B.r_m=\B.r \}).
\EE
The tensor $\C.H_{n}^{\{\a_m\}} (\{\B.r_m\}) $ can be contracted in
various ways. For example, binary contractions $\alpha_1=\alpha_2\,,
\alpha_3=\alpha_4$, {\em etc.}  with $\B.r_1=\B.r_2\,, \B.r_3=\B.r_4$ 
{\em etc.}  produces the correlation functions of dissipation field
$|\B.\nabla T|^2$.

\par~\par
When the ensemble is not isotropic we need to take into account the
angular dependence of $\B.R$, and the scaling behavior consists of
multiple contributions arising from anisotropic effects. The formalism
to describe this is set up in Appendix A and in the forthcoming
Sections.

The correlation functions ${\cal F}_{n}$ satisfy
equation \cite{68Kra}
\begin{eqnarray}
        & & \Big[- \kappa_0 \sum_{i=1}^{n} \nabla^2_i+ 
        \hf \sum_{i,j=1}^{n}
       \kt^{\a\b} (\B.r_i-\B.r_j)\nabla^\a_i\nabla^\b_j
        \Big] 
        {\cal F}_{n}(\{{ \B.r}_{m} \}) 
  \nonumber \\
   &&= 
        \hf \sum_{\{i\ne j\}=1}^{n}
        \Phi({\B.r}_i-{\B.r}_j)
        {\cal F}_{n-2}(\{{\B.r}_m\}_{m\ne i,j}) \,,
\label{bob1} 
\end{eqnarray}
where $\{{\B.r}_{m}\}_{m\ne i,j}$ is the set off all ${\B.r}_{m}$
with $m$ from 1 to $n$, except of $m= i$ and $m= j$. Substituting $
\kt^{\a\b} (\B.r)$ from Eqs.~(\ref{W3}, \ref{eddy-diff})  one
gets:
\begin{eqnarray}
        & & \Big[- \kappa \sum_{i=1}^{n} \nabla^2_i+ 
        \sum_{\{i\ne j\}=1}^{n}
       \C.W^{\a\b} (\B.r_i-\B.r_j)\nabla^\a_i\nabla^\b_j
        \Big]
        {\cal F}_{n}(\{{ \B.r}_{m} \}) 
  \nonumber \\
   &&=\case{1}{2} \sum_{\{i\ne j\}=1}^{n}
        \Phi({\B.r}_i-{\B.r}_j)
        {\cal F}_{n-2}(\{{\B.r}_m\}_{m\ne i,j}) \,,
\label{bob0} 
\end{eqnarray}
where 
\BE\label{kappa}
\kappa  =\kappa _0+ D[\L^\e-\lambda^\epsilon] .
\EE
Here we used that in space homogeneous case 
$ \sum_{i=1}^{n} \B.\nabla_i=0$
and therefore 
$$\Big|\sum_{i=1}^{n} \B.\nabla_i\Big|^2=
\sum_{i=1}^{n}\nabla^2_i+\sum_{\{i\ne j\}=1}^{n}
\nabla^\a_i\nabla^\b_j=0\ .
$$
Consider the $\B.k$--Fourier transform of $\C.F_n$ which is defined as:

\BEA\label{Fourier}
&& (2\pi)^d \delta\Big(\sum_{s=1}^{n}\B.k_s\Big) 
F_{n}(\{{ \B.k}_{m} \}) \\ \nn
&=& \int
\Big[\prod_{m=1}^{n}
d \B.r_{m}\exp (i\B.k_{m} \cdot \B.r_{m})\Big]
 {\cal F}_{n}(\{{ \B.r}_{m} \}) \ .
\EEA
\end{multicols}
\leftline{---------------------------------------------------------------------------}
Here the $\delta$-function applies to a homogeneous ensemble in which
${\cal F}_{n}(\{{ \B.r}_{m}\})$ depends only on {\em differences} of
coordinates.  For $F_{n}(\{{\B.k}_{m}\})$ Eq.~(\ref{bob0}) yields:
\BEA
 && \kappa    F _{n}(\{{ \B.k}_{m} \} )    \sum_{i} k^2_i 
 + 
\frac{\e\, D}{\Od}
\int \limits _{\L ^{-1}}^{\lambda^{-1}}     \frac{  d^d p}{p^{d+\e}} 
P^{\a\b} (\B.p)\sum_{\{i\ne j\}=1}^{n}  k^\a_i k^\b_j 
F_{n}(\B.k_i+\B.p, \B.k_j-\B.p,\{\B.k_{m} \}_{m\ne i,j} )
=  \tilde  \Phi_{n}(\{{ \B.k}_{m} \})\,,
\label{bob3}\\
 \label{tF}
&& \tilde \Phi_{n}(\{{ \B.k}_{m} \})\equiv 
 \frac{(2\pi)^d}{2} \sum_{\{i\ne j\}=1}^{n}
      \Phi(\B.k_i)\delta({\B.k_i+\B.k_j})
  F_{n-2}(\{\B.k_{m} \}_{m\ne i,j}) \ ,\quad \mbox{for~}n > 2 \ ,\\ 
&&\tilde \Phi(\B.k)=\int d\B.R \exp(i\B.k
\cdot\B.R)\Phi(\B.R) \ . \label{defphik}
\EEA
Here $\tilde \Phi(\B.k)$ is the Fourier transform of $\Phi (\B.R)$ and
$\tilde \Phi_2(\B.k)= \Phi(\B.k)$. Equation~(\ref{bob3}) may be
rewritten as:
\BE\label{bob4} 
 F _{n}(\{{ \B.k}_{m} \} ) = -
\frac{\e\, D}{\kappa\,\Od}
\int \limits _{\L ^{-1}}^{\lambda^{-1}}    \frac{  d^d p}{p^{d+\e}} 
P^{\a\b} (\B.p)
\frac{\sum _{\{i\ne j\}=1}^{n}  k^\a_i k^\b_j }
{ \sum_{s=1} ^{n} k^2_s }
F_{n}(\B.k_i+\B.p, \B.k_j-\B.p,\{\B.k_{m} \}_{m\ne i,j} )
+\frac{  \tilde  \Phi_{n}(\{{ \B.k}_{m} \})}
{ \kappa  \sum_{s=1} ^{n}k^2_s }    
\ . 
\EE
This equation will serve as the basis for future analysis in Sect.~\ref{sec:3}.

\section{Scaling of the temperature gradient fields}\label{sec:3}
\subsection{Basic Equations in k-representation}
It appears that Eq.~(\ref{bob4}) is as difficult to solve as
Eq.~(\ref{bob1}).  In fact, very important information about scaling
behavior may be extracted from Eq.~(\ref{bob4}) for small $\e$
\cite{98Spt}. In order to develop our method we will analyze first the
simultaneous, $n$-point correlation functions of the gradient fields
${\cal H}_{n}^{\{\a_m\}} (\{\B.r_m\})$ and $H_{n}^{\{\a_m\}}$ of
Eqs.(\ref{Psi1},\ref{Psi2}): These objects are expressed in terms of
$F_{n}(\{\B.k_{m} \})$ as follows:
\begin{eqnarray}
\label{Psi3}
{\cal H}_{n}^{\{\a_m\}}   (\{\B.r_m\})&=& (2\pi)^{(1-n)d} \int 
\prod_{i=1}^{n}\big[ d^d k_i k^{\a_i}\, 
\exp(i\B.k_i\cdot\B.r_i)\big] F_{n}(\{\B.k_{m} \})\delta
\Big(\sum_{s=1}^{n} \B.k_s\Big), \\
H_{n}^{\{\a_m\}}&=&(2\pi)^{(1-n)d} \int 
\prod_{i=1}^{n}\big [ d^d k_i  k^{\a_i}\,
\big] F_{n}(\{\B.k_{m} \}) \delta
\Big(\sum_{s=1}^{n} \B.k_s\Big)\ . \label{Psi4}
\end{eqnarray}
From this and Eq.~(\ref{bob4}) one gets:
\begin{eqnarray} \label{bob5} 
 H_{n}^{\{\a_m\}}&=&
   - \frac{\e\, D}{\kappa\,\Od}
\int\frac{\prod_{s=1}^{n} k_s ^{\a_s}d^d k_s}
{(2\pi)^{(n-1)d}}\delta\Big(\sum_{s=1}^{n}
\B.k_s\Big) \int \limits _{\L ^{-1}}
^{\lambda^{-1}}     \frac{  d^d p}{p^{d+\e}} 
\frac{P^{\a\b} (\B.p) \sum_{\{i\ne j\}=1}^{n}  k^\a_i k^\b_j }
{ \sum_{s=1} ^{n} k^2_s }F_{n}(\B.k_i
+\B.p, \B.k_j-\B.p,\{\B.k_{m} \}_{m\ne i,j} )\\ \label{bob6}
&& + \Psi_{n}^{\{\a_m\}}\ ,\qquad
\Psi_{n}^{\{\a_m\}}\equiv 
\int\frac{\prod_{s=1}^{n} k_s ^{\a_s}d^d k_s }{(2\pi)^{(n-1)d}}
\frac{  \tilde  \Phi_{n}(\{{ \B.k}_{m} \})}
{ \kappa  \sum_{s=1}^ {n}k^2_s} \delta\Big(\sum_{s=1}^{n}
\B.k_s\Big) \ .   
\EEA
Shifting the dummy variables $\B.k_i-\B.p\to\B.k_i$ and
$\B.k_j+\B.p\to\B.k_j$ we have another representation of this
equation:
\begin{eqnarray}
H_{n}^{\{\a_m\}} &=&
   -  \frac{\e\, D}{\kappa\,\Od}
\int\frac{ \prod_{s=1}^{n} d^d  k_s }{(2\pi)^{(n-1)d}}
\delta\Big(\sum_{s=1}^{n}
\B.k_s\Big)   \sum_{\{i\ne j\}=1}^{n}  
\int \limits _{\L ^{-1}}^{\lambda^{-1}}   
  \frac{  d^d p}{p^{d+\e}} 
\frac{ (k_i^{\a_i}-p^{\a_i}) (k_j^{\a_j}+p^{\a_j}) 
P^{\a\b} (\B.p)   k^\a_i k^\b_j  }
{ 2p^2 +2\B.p\cdot(\B.k_j-\B.k_i)
+ \sum_{s=1} ^{n} k^2_s }
\nonumber\\&\times&\prod_{s=1, s\ne i,j }^{n} 
  k_s ^{\a_s}F_{n}(\{\B.k_{m} \})
+ \Psi_{n}^{\{\a_m\}}\ .\label{bob7} 
\end{eqnarray}
In order to analyze this equation further we choose to
nondimensionalize all the wave vectors by $\Lambda$. We write $\tilde
\B.k_s=\Lambda \B.k_s$, $\tilde \B.p=\Lambda \B.p$ etc, and for
simplicity drop the tilde signs at the end. We simplify the appearance
of the equation further by introducing the definition of the
dimensionless function
\begin{equation}
A^{\alpha_i\alpha_j}_{\beta_i\beta_j}(\{k^2_m\}_{s\ne
i,j}\B.k_i,\B.k_j,\B.p) =-\int \frac{d\hat\B.p}{\Omega_d} \frac{
(k_i^{\a_i}-p^{\a_i}) (k_j^{\a_j}+p^{\a_j}) P^{\beta_i\beta_j}
(\hat\B.p)} { 2p^2 +2\B.p\cdot(\B.k_j-\B.k_i)
+ \sum_{s=1} ^{n} k^2_s
} \ , \label{defA}
\end{equation}
where  $d\hat \B.p$ stands for integrating over all
the angles of the unite vector $\hat\B.p\equiv \B.p/p$. The  resulting
equation is
\begin{eqnarray}
H_{n}^{\{\a_m\}} &=& g
\int\frac{ \prod_{s=1}^{n} d^d  k_s }{(2\pi)^{(n-1)d}}
\delta\Big(\sum_{s=1}^{n}
\B.k_s\Big)   \sum_{\{i\ne j\}=1}^{n}  
\int \limits _1^{\Lambda/\lambda}  
  \frac{ \epsilon d p}{p^{1+\e}} 
A^{\alpha_i\alpha_j}_{\beta_i\beta_j}(\{k^2_m\}
_{s\ne i,j}\B.k_i,\B.k_j,\B.p) k^{\b_i}
k^{\b_j}  \nonumber\\&\times&\prod_{s=1, s\ne i,j }^{n}   \!\!  k_s
^{\a_s}F_{n}(\{\B.k_{m} \}) + \Psi_{n}
^{\{\a_m\}}\ , \label{HA} 
\end{eqnarray}
\rightline{--------------------------------------------------------------------------}
\begin{multicols}{2}
where the dimensionless factor $g$ is 
\begin{equation}
g\equiv \frac{D\Lambda^\epsilon}{\kappa_0
+D(\Lambda^\epsilon-\lambda^\epsilon)} \
. \label{g}
\end{equation}
In fact, we should recognize that the natural expansion parameter is
actually not $g$, but $\tilde g$, where
\begin{equation}
\tilde g\equiv
g\int_1^{\Lambda
/\lambda}\frac{\epsilon dp}{p^{1+\epsilon}} \ .
\end{equation}
Evaluating the integral we find
\begin{equation}
\tilde g =
\frac{D[\Lambda^\epsilon-\lambda^\epsilon]}{\kappa_0
+D(\Lambda^\epsilon-\lambda^\epsilon)}
\ .
\end{equation}
We will see below that $\tilde g$ can take on very different values in
different limiting cases. In particular it can be of $O(\epsilon)$ or of $O(1)$
depending on the order of limits. The relevant limit for the physics
at hand will be discussed below. At this point we
perform a calculation of $H_{n}^{\{\a_m\}}$ to first order in
$\tilde g$ in all the  sectors of the symmetry group. 
\subsection{Theory to first order in $\tilde g$}
The theory for $F_{n}$ and $\B.H_{n}$ can be formulated
iteratively, resulting in the following series:
\BE\label{ser1}
F_{n}(\{\B.r_m\})
=\sum_{q=0}^\infty F_{n,q}(\{\B.r_m\}) \,,\quad
H_{n}^{\{\a_m\}}=\sum_{q=0}^\infty H_{n,q}^{\{\a_m\}}
\ .
\EE
Here $ F_{n,q}$ is the result of the $q$-step of the iteration
procedure of Eq.~(\ref{bob4})$, F_{n,q}\propto {\tilde g}^q$.  There are
two contributions for each term of order ${\tilde g}^q$.  One arises
from substituting $ F_{n,q-1}$ into the integral on the RHS of
(\ref{bob4}), and the second arises from $F_{n-2,q}$ which appears in
$\tilde \Phi_{n}$ according to (\ref{tF}). Correspondingly, also
$H_{n,q}$ has two contributions. One is obtained from Eq.~(\ref{HA})
when we substitute $F_{n,q-1}$ in the integral on the RHS, and the
second when we substitute $F_{n-2,q}$ in the term denoted as
$\Psi_{n}$.

In this Sect. we compute explicitly $H_{n,1}$. Analyzing the
relative importance of these two contributions to $H_{n,1}$ we found
that the 2nd contribution is negligible compared to the 1st when
$\Lambda \ll L$. In other words, we can disregard the contribution to
$H_{n,1}$ which arises from $F_{n-2,q}$. This means that for the
sake of our iteration procedure we can replace $\tilde\Phi_{n}$ in
(\ref{bob4}) by the quantity $\tilde \Phi_{n,0}$ which is of
$O(\tilde g^0)$.  This means that instead of Eq.~(\ref{bob4}) we
iterate
\end{multicols}
\leftline{----------------------------------------------------------------------------}
\BEA\label{bob40} 
 F _{n}(\{{ \B.k}_{m} \} )&=& -
\frac{\e\, D}{\kappa\,\Od}
\int \limits _{\L ^{-1}}^{\lambda^{-1}}    \frac{  d^d p}{p^{d+\e}} 
P^{\a\b} (\B.p)
\frac{\sum _{\{i\ne j\}=1}^{n}  k^\a_i k^\b_j }
{ \sum_{s=1} ^{n} k^2_s }
F_{n}(\B.k_i+\B.p, \B.k_j-\B.p,\{\B.k_{m} \}_{m\ne i,j} )
+F_{n,0}(\{{ \B.k}_{m} \})\,, \\  \label{last}
F_{n,0}(\{{ \B.k}_{m} \})&=&\frac{  \tilde  \Phi_{n,0}(\{{ \B.k}_{m} \})}
{ \kappa  \sum_{s=1} ^{n}k^2_s }    
\,, \qquad 
\tilde \Phi_{n,0}(\{{ \B.k}_{m} \})\equiv
 \frac{(2\pi)^d}{2} \sum_{\{i\ne j\}=1}^{n}
      \Phi(\B.k_i)\delta({\B.k_i+\B.k_j})
  F_{n-2,0}(\{\B.k_{m} \}_{m\ne i,j})  \  .
\EEA
 Thus we are interested in calculating
\FL
\BE   \label{H1F0} 
H_{n,1}^{\{\a_m\}} = \tilde g
\int\frac{ \prod_{s=1}^{n} d^d  k_s }{(2\pi)^{(n-1)d}}
 \delta\Big(\sum_{s=1}^{n}
\B.k_s\Big) \!\!\!  \sum_{\{i\ne j\}=1}^{n} 
\int \limits _1^{\Lambda/\lambda} 
   \frac{ \epsilon d p}{p^{1+\e}} 
A^{\alpha_i\alpha_j}_{\beta_i\beta_j}
(\{k^2_m\}_{s\ne i,j}\B.k_i,\B.k_j,\B.p) k^{\b_i}
k^{\b_j} \!\!\!  \prod_{s=1, s\ne i,j }^{n}\!\!\!   k_s
^{\a_s}F_{n,0}(\{\B.k_{m} \})\ .
\EE
Recall that the function $\Phi(\B.k)$ is constant for $kL\ll 1$ and it
vanishes for $kL\gg 1$. Therefore the leading contribution to the integrals 
(\ref{H1F0}) over  $\B.k_i$ comes from the region $k_i L\leq 1$. 
In integral~(\ref{H1F0}) $p>1/ \L$ and in all our approach we
consider $L\gg \L$. Therefore in (\ref{H1F0}) $p\gg k_j$ and this
equation may be simplified up to:
\BE
H_{n,1}^{\{\a_m\}} = \tilde g
\int\frac{ \prod_{s=1}^{n} d^d  k_s }{(2\pi)^{(n-1)d}}
\delta\Big (\sum_{s=1}^{n}
\B.k_s\Big)   \sum_{\{i\ne j\}=1}^{n}  
\int \limits _1^{\Lambda/\lambda}    \frac{ \epsilon d p}{p^{1+\e}} 
A^{\alpha_i\alpha_j}_{\beta_i\beta_j} k^{\b_i}
k^{\b_j}  \prod_{s=1, s\ne i,j }^{n}   \!\!  k_s
^{\a_s}F_{n,0}(\{\B.k_{m} \})\ , \label{H1F1} 
\EE
\begin{multicols}{2}
where now
\BE\label{int1}
A^{\a_i\a_j}_{\b_i\b_j}\equiv \frac{1}{2 \Od}\int d \hat \B.p \, 
\hat p^{\a_i}  \hat p^{\a_j}  P_{\b_i\b_j} (\hat \B.p)\,,
\EE
is the constant tensor that obtains from the tensor function
(\ref{defA}) when all $k_s\ll p$. Performing the all the
wave-vector integrals we observe that the explicit $\epsilon$ is
canceled by integral over $p$. Accordingly 
\BE 
H_{n,1}^{\{\a_m\}} = \tilde g
\sum_{\{i\ne j\}=1}^{n}A^{\a_i\a_j}_{\b_i\b_j}
H _{n,0}^{\b_i\b_j\{\a_{m} \}_{m\ne i,j}}\ .
\label{per2} 
\EE 
An actual integration in~(\ref{int1}) yields
\BE\label{A}
A^{\a_i\a_j}_{\b_i\b_j}=\frac{\d_{\a_i\a_j}\d_{\b_i\b_j}(d+1)-
 \d_{\a_i\b_i }\d_{\a_j\b_j}-   \d_{\a_i\b_j }\d_{\a_j\b_i} 
}{2(d+2)(d-1)}\ .
\EE
\subsection{Analysis in all the anisotropic sectors}
We note that Eqs.~(\ref{bob0}) contain only isotropic operators. On
the other hand $\Phi(\B.r_i-\B.r_j)$ can be anisotropic, depending on
the direction of the vector $\B.r_i-\B.r_j$. Since he equations are
{\em linear}, we can expand all the objects in terms of the
irreducible representations of the SO($d$) group of all rotations, and
be guaranteed that the solutions foliate in the sense that different
irreducible representations cannot be mixed.  This considerations are
valid for all the equations in this theory, including
Eq.~(\ref{per2}). To know which irreducible representations we need to
use in every case one has to consult Appendix A. After doing so one
notes that for any order $q$, the tensors $ H^{\{\alpha_m\}}_{n,q}$
are constant tensors, fully symmetric in all their indices. Using the
exposition of Appendix A we know that the projections on the
irreducible representations of the SO($d$) symmetry group must be of
the form
\begin{equation}
 H^{\{\alpha_m\}}_{n,q,\ell}=\lambda_q^{(\ell)}
B^{\{\alpha_m\}}_{\ell,\sigma,n}
\ . \label{PsiB}
\end{equation}
Our first order calculation is aimed at finding the ratio
$\lambda_1^{(\ell)}/ \lambda_0^{(\ell)}$. Substituting
(\ref{A},\,\ref{PsiB}) in (\ref{per2}) we find
\begin{eqnarray}
&&H^{\{\alpha_m\}}_{n,1,\ell}=\frac{\tilde g}{2(d+2)(d-1)}
\Big[(d+1)
\sum_{i\ne j}\delta^{\alpha_i\alpha_j}
\delta_{\beta_i\beta_j}\lambda_0^{(\ell)}\nonumber \\  
&\!\times\!\!& \! B^{\beta_i\beta_j\{\alpha_m\}
m\ne i,j}_{\ell,\sigma,n}\!\!-\!\sum_{i\ne j}[
\delta^{\alpha_i}_{~\beta_i}\delta^{\beta_i}_{~\beta_j}
+\delta^{\alpha_i}_{~\beta_j}
\delta^{\alpha_j}_{~\beta_i}]\lambda_0^{(\ell)}
B^{\beta_i\beta_j\{\alpha_m\}}_
{\ell,\sigma,n}\!\Big]\nonumber\\
&=&\frac{\tilde g}{(d+2)(d-1)}
\Big[\frac{(d+1)z_{n,\ell}}{2}-n(n-1)
\Big]\lambda_0^{(\ell)}
B^{\{\alpha_m\}}_{\ell,\sigma,n} \ . \nn
\end{eqnarray}
Defining now $A_{n}^{(\ell)}$ via the relation
\BE\label{exp1}
\B.H_{n,1}^{(\ell)}=\tilde gA_{n}^{(\ell)}
\B.H^{(\ell)}_{n,0}\,,
\EE
we conclude that
\BE\label{exp4}
A^{(\ell ) }_{n }= \frac{2n(d+n)}{d+2} -\frac{(d+1)
\ell (\ell + d -2)}   {2(d-1)(d+2)} \ .
\EE

\subsection{Interpretation of the result}
To interpret the result (\ref{exp1}-\ref{exp4}) we should observe that
the nature of the theory that we develop depends on the order of the
limits that we take. We should recognize that the quantity
$H_{n}^{\{\a_m\}}$ does not exist without an inner (ultra-violet)
cutoff. We are thus interested in limiting values of $\tilde g$
subject to the condition that $\eta$ is finite. Thus one order of
limits that makes sense is $\lambda\to 0$ first ( corresponding to the
Reynolds number going to infinity first), and then $\epsilon$ going to
zero second, but keeping $\eta$ fixed [for example by controlling
$\kappa_0$ in Eq.~(\ref{eta})]. Another order of limits is $\epsilon\to
0$ first, (still keeping $\eta$ fixed, but very small) with $\lambda$
being fixed and larger than $\eta$.

If we have $\lambda\to 0$ first, and then when $\epsilon\to 0$ second
we find that the expansion parameter is close to unity:
\begin{equation}
\tilde g \approx 1-\left(\frac{\eta}{\Lambda}\right)
^\epsilon \ , \quad \mbox{for~~}
\epsilon\ln\left(\frac{\eta}{\Lambda}\right) \ll 1 \ . \label{lamfirst}
\end{equation}
Thus we cannot stop at (\ref{exp1}), and we are forced to consider
higher order terms in the expansion in $\tilde g$ and appropriate
resummations. This is done in Subsect. F.  On the other hand, if
$\epsilon\to 0$ first, we find an apparently ``small" expansion
parameter that is proportional to $\epsilon$:
\begin{equation}
\tilde g\approx  \epsilon\ln\left(\frac{\Lambda}
{\lambda}\right) \ , \quad \mbox{for~~}
\epsilon\ln\left(\frac{\eta}{\Lambda}\right) \gg 1 \ . \label{epsfirst}
\end{equation}
\subsection
{Exponentiating using Renormalization Group Equations}\label{RG}
Using Eq. (\ref{epsfirst}) in Eq.~(\ref{exp1}) we get:
\begin{equation}
\B.H_{n}^{(\ell)}=\left[1+\epsilon
A_{n}^{(\ell)}\ln\left(\frac{\Lambda}{\lambda}
\right)+O(\epsilon^2)\right]
\B.H^{(\ell)}_{n,0}  \ . \label{firstord}
\end{equation}
If we expect that $\B.H_{n}^{(\ell)}$ is a scale invariant function
of $\Lambda/\lambda$ we can interpret Eq.~(\ref{firstord}) as the
beginning of an expansion that can be re-summed into a power law
\begin{equation}
\B.H_{n}^{(\ell)}=\left(\frac{\Lambda}{\lambda}\right)^{\epsilon
A_{n}^{(\ell)}}\B.H^{(\ell)}_{n,0}  \ . \label{expon}
\end{equation}
Of course, this is hardly justified just by examining the
$O(\epsilon)$ term, since one can have more than one branch of scaling
exponents proportional to $\epsilon$. If we have $m$ branches only the
analysis up to $O(\epsilon^m)$ can reveal this. We return to this issue
in the next Subsection. An additional issue is the magnitude of
$\lambda$ that can be arbitrarily small, making any re-exponentiation
even more dubious. To overcome this problem one usually invokes the
renormalization group equations to justify the exponentiation.  We
shortly present this method next. In doing so we want to argue that
for the case in question there is nothing more in this approach than direct
re-exponentiation as long as higher order in $\epsilon$ are not included.

Within the renormalization group method \cite{98Spt} one considers
Eq.~(\ref{firstord}) as the ``bare" value of $\B.H_{n}^{(\ell)}$,
$\B.H_{n,B}^{(\ell)}$.  One then seeks a multiplicative
renormalization group by defining a renormalized function
\BE
\B.H_{n,R}^{(\ell)}(\mu,\L,\dots)=Z_H(\mu,\Lambda,\lambda)
\B.H_{n,B}^{(\ell)}(\Lambda,\lambda) \ . \label{renor}
\EE
Here $\mu$ is a ``running length", and the only $\mu$ dependence of the
RHS is through the $Z_H$ function. Defining $Z_H$ so that it eliminates 
the dependence of the LHS on $\lambda$ and setting the initial condition
$\B.H_{n,R}(\L,\L,\dots)=\B.H^{(\ell)}_{n,0}$ we  get:
\BE
Z_H(\mu,\lambda)=1+\e A_{n}^{(\ell)}\ln(\lambda/\mu) \ . \label{ZH}
\EE
From equation (\ref{renor}) we get:
\BE\label{dieq}
\frac{d\ln(\B.H_{n,R}^{(\ell)}(\mu,\L,\dots))}{d \ln\mu}=
\gamma_{_H} \ ,
\EE
where
\BE
\gamma_{_H}=-\e A_{n}^{(\ell)}+O(\e^2) \ .
\EE
Solving the differential equation (\ref{dieq}) we get:
\BE
\B.H_{n,R}^{(\ell)}(\mu,\L,\dots)=\left(\frac{\L}{\mu}\right)^{\e
A_{n}^{(\ell)}}\B.H_{n,0} \ .
\EE
Exponentiating (\ref{ZH}) and solving Eq. (\ref{renor}) in favor of
$\B.H_{n,B}^{(\ell)}$ we recover Eq. (\ref{expon}). Note that 
the inner scale in this case is $\lambda$, since $\lambda>\eta$.
This fact casts an additional doubt on this limit of the theory, since
it misses altogether the existence of the Batchelor regime \cite{53Bet}
between $\lambda$ and $\eta$. For all these reason we tend to disqualify
(\ref{expon}) despite the relative simplicity of its derivation.
We turn next to the other limiting procedure.

\subsection{Theory for $\lambda\to 0$ first} 

Considering the limit $\lambda\to 0$ first, Eq.~(\ref{exp1}) is still
valid, but now $\tilde g$ is of order unity, and we cannot justify
re-exponentiation by any stretch of the imagination.

\narrowtext
\begin{figure}
\epsfxsize=5truecm
\epsfbox{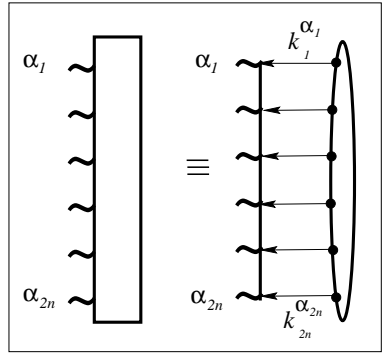}
\caption{The graphic representation of
Eq.~(3.3)(Psi4).}
\label{Fig1} 
\end{figure}
\begin{figure}
\epsfxsize=8.5truecm
\epsfbox{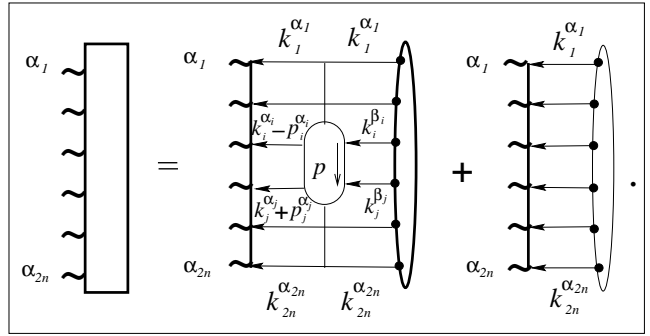}
\caption{The graphic representation of
Eq.~(3.8)(HA).}
\label{Fig2} 
\end{figure}

\noindent
The small parameter $\epsilon$ seems to have disappeared. This forces
us to consider all the higher order terms in $\tilde g$ to understand
how to resum them. We will see that at the end $\epsilon$ reappears.

The re-summation of the $\tilde g$ dependence is aided significantly by
some graphic representations of the relevant equations and their
perturbative solutions. In Fig.~\ref{Fig1}  we represent graphically the
definition (\ref{Psi4}) of $H_{n}^{\{\alpha_m\}}$ in terms of
$F_{n}(\{\B.k_m\})$. This helps us to introduce the basic
diagrammatic notations.

A solid long rectangle with $n$ $\alpha_m$ wavy lines stands for
$H_{n}^{\{\a_m\}}$, while the elongated solid ellipse represents
$F_{n}$. The $\B.k_s$ wave-vectors are denoted by arrows, and the dots
represent multiplications. The vertical line connecting all the arrow
heads stands for the integration with a delta function over the sum of
all the $\B.k$ vectors. Fig.~\ref{Fig2} is a graphic notation of
Eq.~(\ref{HA}). The little ellipse with a vertical arrow designated by
$\B.p$ stands for the tensor function
$A^{\alpha_i\alpha_j}_{\beta_i\beta_j}(\{\B.k_s\}_{s\ne
i,j}\B.k_i,\B.k_j,\B.p)$. This ellipse is involved in the integration
over the vector $\B.p$ in the loop to the left of the ellipse with a
weight consisting of the sum of squares of the $\B.k$ vectors.

The thin elongated ellipse in the second term on the RHS stands for
the zero'th order term $F_{n,0}$, cf. Eq.~(\ref{last}). The actual
values of the wave-vectors are indicated in this diagram. In later
diagrams, Fig.3, we drop this obvious notation.  In Fig.~\ref{Fig3} we
display the perturbative solution which results from the iteration
procedure in Eq.~(\ref{bob40}), the result of which is substituted in
Eq.~(\ref{bob5}).

\begin{figure}
\epsfxsize=8.5truecm
\epsfbox{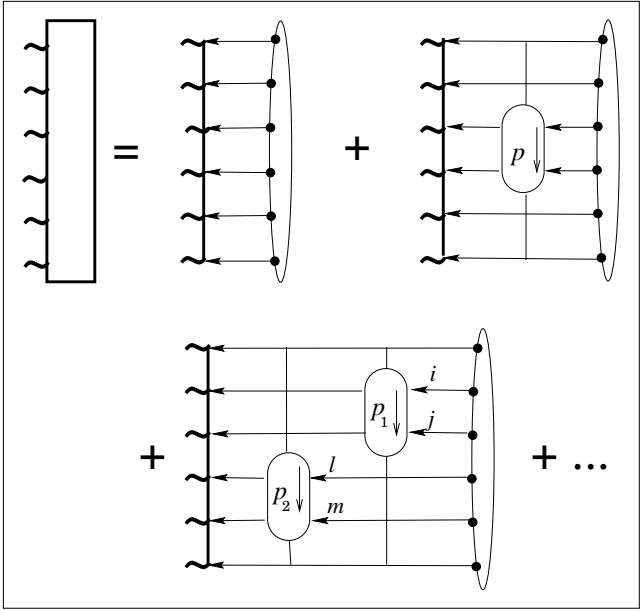}
\caption{The graphic representation of
the iteration scheme}
\label{Fig3} 
\end{figure}

The first diagram on the RHS is $\B.H^{(\ell)}_{n,0}$, whereas the
second is the first order term $\B.H^{(\ell)}_{n,1}$,
Eq.~(\ref{exp1}). The last diagram shown is $\B.H^{(\ell)}_{n,2}$.
One should note that when the $A^{\alpha_i\alpha_j}_{\beta_i\beta_j}$
ellipse appears once we have a sum over all pairs $i,j$ indices with
$i\ne j$. When it appears twice there is a double sum, with respect to
the pairs $i\ne j$ and $\ell\ne m$.  In analyzing such diagrams one
needs to identify three distinct possibilities. These are denoted as
case (a), $i=\ell$, $j=m$, case (b) $i=\ell$ but $j\ne m$, and case
(c) where all the indices are different. There are two integrals over
$\B.p_1$ and $\B.p_2$ in the loops to the left of the corresponding
ellipse. We refer to the functions $\B.A$ of Eq. (\ref{defA}) which
appears in these integrals as $\B.A_1$ and $\B.A_2$
respectively. Analyzing the integrals it is useful to separate the
discussion to region I in which $p_1>p_2$, and region II in which
$p_2>p_1$. In region I we can neglect the contribution of $\B.p_2$ and
all $\B.k_s$ with respect to $\B.p_1$. Accordingly we have two
independent integrals and sums, and the result is therefore
\begin{equation}
\B.H_{n,2}^{(\ell)}=\case{1}{2}[\tilde gA_{n}^{(\ell)}]^2
\B.H^{(\ell)}_{n,0}\ , \quad \mbox {(region I)} \ ,
\end{equation}
where the factor 1/2 stems from the fact that the volume of region I
is a half of the whole volume of $(\B.p_1,\B.p_2)$-space. In region II
we should distinguish between the cases (a), (b) and (c), for which
the evaluation of $\B.A_2$ will be different. In case (c)
Eq.~(\ref{defA}) shows that $\B.A_2$ is of the order of $p_2^2/p_1^2$,
which is small. In case (b) $\B.A_2$ is of the order of $p_2/p_1$,
which is still small. Only case (a), in which the the loops appear as
two rungs on the same ladder, we have $\B.A_2$ of the order of unity.
The actual calculation of this integral is presented in Appendix B,
with the final result

\begin{equation}
\B.H_{n,2}^{(\ell)}=\case{1}{2}\tilde g^2A_{n}^{(\ell)}
\B.H^{(\ell)}_{n,0}\ , \quad \mbox {[region II, case (a)]}\ . 
\label{casea}
\end{equation}
Together the second order result for $\B.H_{n,2}^{(\ell)}$ is 
\begin{equation}
\B.H_{n,2}^{(\ell)}=\case{1}{2}\tilde g^2 
\Big[A_{n}^{(\ell)}+\left(
A_{n}^{(\ell)}\right)^2\Big] \B.H_{n,0}^{(\ell)}\ . \label{second}
\end{equation}

Our aim is to find the fully resummed form, correct to all order in
$\tilde g$ and $A_{n}^{(\ell)}$, of $\B.H_{n}^{(\ell)}$. We can
express it in the form
\begin{equation}
\B.H_{n}^{(\ell)}\equiv K(\tilde g,A_{n}^{(\ell)})
\B.H_{n,0}^{(\ell)} \ , 
\label{defK}
\end{equation}
where the function $ K(\tilde g,A_{n}^{(\ell)})$ is
represented as the double infinite sum
\begin{equation}
 K( \tilde g,A)=1+\sum_{m=1}^\infty 
A^m\sum_{s=m}^\infty D_{m,s} \tilde g^s \ . \label{Ksum}
\end{equation}
Up to now we have information about $D_{1,1}=1$ and
$D_{1,2}=D_{2,2}=\case{1}{2}$.

In Appendix C we derive the following recurrent relation for the
higher order terms
\BE\label{D1s}
D_{1,s}=\frac{1}{s} \ , \quad
D_{m+1,s}=\frac{1}{s}\sum_{q=m}^{s-1} D_{q,m} \ .
\end{equation}
Using (\ref{D1s}) in Eq.~(\ref{Ksum}) we find the contribution
proportional to $A$:
\begin{equation}
K_{1}( \tilde g,A)=A\sum_{s=1}^\infty 
\frac{\tilde g^s}{s}=-A\ln(1- \tilde g) \ . \label{K2s1}
\end{equation}
Considering all the terms quadratic in A, and using the recurrent
relations to determine $D_{2,s}$ we find
\begin{equation}
K_{2} ( \tilde g,A)=A^2\sum_{s=2}^\infty \frac{ \tilde g^s}{s}
\sum_{q=1}^{s-1}\frac{1}{q}
\ . \label{Kn2}
\end{equation}
This double sum is computed in Appendix C with the result
\begin{equation}
K_{2}( \tilde g,A)=\frac {1}{2} [-A\ln(1- \tilde g)]^2 
\ . \label{Kn2final}
\end{equation}
The general result can be derived using similar techniques with the
result
\begin{equation}
K_{m}( \tilde g,A)=\frac {1}{m!} [-A\ln(1- \tilde g)]^m \ . 
\label{Kmfinal}
\end{equation}
Accordingly we conclude with the series for $K( \tilde g,A)$:
\begin{equation}
K( \tilde g,A) =\sum_{m=0}^\infty \frac{[-A
\ln(1- \tilde g)]^m}{m!} = \exp[-A\ln(1- \tilde g)] \ . \label{Kfinal}
\end{equation} 
Using  now Eq.~(\ref{lamfirst}) 
one finds
\begin{equation} 
K ( \tilde g,A) =
\left(\frac{\Lambda}{\eta}\right)^{\epsilon A} \ . \label{Knexp}
\end{equation}
Finally, using Eq. (\ref{defK}) we have the final result
\begin{equation}
\B.H_{n}^{(\ell)} = \left(\frac{\Lambda}{\eta}\right)^{\epsilon
A_{n}^{(\ell)}}\B.H_{n,0}^{(\ell)} \ . \label{Hfinal}
\end{equation}
We are pleased to find that the inner scale is now $\eta$, in
agreement with our expectation. The exponentiation was achieved 
naturally in the present case. 

In assessing this result, we need to return to a delicate point in the
derivation of Eq.~(\ref{Hfinal}). The procedure involved summing all
the terms of the order of unity, (powers of $\tilde g\approx 1$),
while neglecting terms of $O(\epsilon)$. However, the sums
(\ref{Kn2final}-\ref{Kmfinal}) result in expressions containing
$(\tilde g-1)\propto \epsilon$. In other words, we end up with terms
that appear of the same order as those neglected during the
procedure. In order to justify the results (\ref{Hfinal}) we must go
back and analyze contributions of $O(\epsilon)$. These appear for
example in contributions in which the ``rungs" appear on adjacent
ladders, like case (b) in region II. The two ``simple" rungs appearing
in two adjacent ladders can be now considered as a single compounded
rung.  We focus now on the infinite set of diagrams in which this
compounded rung repeats many times. The sum of such diagrams will again
generate results containing $(\tilde g-1)$, and in the end will be
responsible for terms of $O(\epsilon^2)$ in the scaling exponent.

The reason for this phenomenon is the structure of the iterative solution.
Sums of terms of order $\tilde g$ have cancellations, leading eventually
to a result of $O(\epsilon)$. The sum of terms of $O(\epsilon)$ have a
very similar structure, just we a redefined ``rung". Therefore they
automatically generate another factor of $\epsilon$ by re-summation. This
phenomenon repeats in higher orders, again by redefining what do mean
by a ``rung".

Thus Eq.~(\ref{Hfinal}) can be considered as the final result for the
scaling of the fused correlation function of gradient fields with the
exponent correct to $O(\epsilon)$. In the next section we turn to the
calculation of the scaling exponent of the unfused correlation
function of the scalar field itself. We will show that the exponents
computed in both methods agree when the same objects are
evaluated. This agreement is connected in the final section with the
existence fusion rules that control the asymptotic properties of
unfused correlation functions when some coordinates are fused
together.

\section{zero modes in the anisotropic sectors}
\subsection{Calculation of the correlation functions}
In this section we consider the zero-modes of Eq.~(\ref{bob1}). In other
words we seek solutions $Z_{n}(\{\B.r_m\})$ which in the inertial interval
solve the homogeneous equation
\begin{equation}
\sum_{i\ne j=1}^{n}
       \kt^{\a\b} (\B.r_i-\B.r_j)\nabla^\a_i\nabla^\b_j
        Z_{n}(\{{ \B.r}_{m} \}) =0 \ . \label{zeromodes}
\end{equation}
We allow anisotropy on the large scales. Since all the operators here
are isotropic and the equation is linear, the solution space foliate
into sectors $\{\ell,\sigma\}$ corresponding the the irreducible
representations of the SO($d$) symmetry group. Accordingly we write the
wanted solution in the form
\begin{equation}
 Z_{n}(\{{ \B.r}_{m} \})= \sum_{\ell,\sigma}
Z_{n,\ell,\sigma}(\{{ \B.r}_{m} \})
\ , \label{Zn}
\end{equation}
where $Z_{n,\ell,\sigma}$ is composed of  functions which transform
according to the  ($\ell,\sigma$)- irreducible
representations of SO($d$).  Each of these
components is now expanded in $\epsilon$. In other words, we write,
in the notation of Ref.\cite{96BGK},
\begin{equation}
Z_{n,\ell,\sigma} =E_{n,\ell,\sigma}+\epsilon G_{n,\ell,\sigma}
+O(\epsilon^2) \ .
\end{equation}
For $\epsilon=0$ Eq.~(\ref{zeromodes}) simplifies to
\begin{equation}
\sum_{i=1}^{n} \nabla_i^2 E_{n,\ell,\sigma}
(\{{ \B.r}_{m} \}) =0 \ , \label{EqE}
\end{equation}
for any value of $\ell,\sigma$. Next we expand the operator in
Eq.~(\ref{zeromodes}) in $\epsilon$ and collect the terms of
$O(\epsilon)$:
\begin{equation}
\sum_{i=1}^{n} \nabla_i^2 G_{n,\ell,\sigma}(\{{ \B.r}_{m}
\})=V_{n}E_{n,\ell,\sigma}(\{{ \B.r}_{m} \}) \ , \label{firsto}
\end{equation}
where $\epsilon V_{n}$ is the first order term in the expansion of the
operator in (\ref{zeromodes}):
\begin{equation}
V_{n} \equiv \sum_{j\ne
k=1}^{n}\Big[\delta^{\alpha\beta}\ln(r_{jk})-{r_{jk}^\alpha
r_{jk}^\beta\over (d-1)r_{jk}^2}\Big]\nabla_j^\alpha\nabla_k^\beta \ ,
\label{defVn}
\end{equation}
where $\B.r_{jk}\equiv \B.r_j-\B.r_k$.

In solving Eq.~(\ref{EqE}) we are led by the following considerations:
we want scale invariant solutions, which are powers of $\B.r_{jk}$. We
want analytic solutions, and thus we are limited to
polynomials. Finally we want solutions that involve all the $n$
coordinates for the function $E_{n,\ell,\sigma}$; solutions with fewer
coordinates do not contribute to the structure functions
(\ref{Sunfused}).  To see this note that the structure function is a
linear combination of correlation functions. This linear combination
can be represented in terms of the difference operator
$\delta_j(\B.r,\B.r')$ defined by:
\begin{equation}
\delta_j(\B.r,\B.r') {\cal F}(\{\B.r_m\})\equiv {\cal
F}(\{\B.r_m\})|_{\B.r_j=\B.r}
-{\cal F}(\{\B.r_m\})|_{\B.r_j=\B.r'} \ . \label{diffop}
\end{equation}
Then, 
\begin{equation}
S_{n}(\B.r_1,\B.r_2\dots \B.r_{n}\B.r'_{n}) = \prod_j
\delta_j(\B.r_j,\B.r'_j) {\cal F}(\{\B.r_m\}) \ . \label{SinF}
\end{equation}
Accordingly, if ${\cal F}(\{\B.r_m\})$ does not depend on $\B.r_k$,
then $\delta_k (\B.r_k,\B.r'_k){\cal F}(\{\B.r_m\})=0$
identically. Since the difference operators commute, we can have no
contribution to the structure functions from parts of ${\cal F}$ that
depend on less than $n$ coordinates. Finally we want the minimal
polynomial because higher order ones are negligible in the limit
$r_{jk}\ll \Lambda$. Accordingly, $E_{n,\ell,\sigma}$ with $\ell\le n$
is a polynomial of order $n$.  Consulting Appendix A for the
irreducible representations of the SO($d$) symmetry group, we can
write the most general form of $E_{n,\ell,\sigma}$, up to an arbitrary
factor, as
\begin{equation} E_{n,\ell,\sigma}=r_1^{\alpha_1}\dots
r_{n}^{\alpha_{n}} B^{\alpha_1\dots \alpha_{n}}_{n,\ell,\sigma}
+[\dots] \ ,
\end{equation}
where $[\dots]$ stands for all the terms that contain less than $n$
coordinates; these do not appear in the structure functions, but
maintain the translational invariance of our quantities. The
appearance of the tensor $B^{\alpha_1\dots \alpha_n}_{n,\ell,\sigma}$
of Appendix A is justified by the fact that $E_{n,\ell,\sigma}$ must
be symmetric to permutations of any pair of coordinates on the one
hand, and it has to belong to the $\ell,\sigma$ sector on the other
hand. This requires the appearance of the fully symmetric tensor
(\ref{Birr}).

In light of Eqs.~(\ref{firsto}-\ref{defVn}) we seek solution for
$G^{(\ell)}_{n} (\{{ \B.r}_{m}\})$ of the form
\begin{equation}
G_{n,\ell\sigma}(\{{ \B.r}_{m}\})=\sum_{j\ne
k}H^{jk}_{\ell,\sigma}(\{{ \B.r}_{m}\}) \ln(r_{jk})
+H_{\ell,\sigma}(\{{ \B.r}_{m}\}) \ , \label{ansatzG}
\end{equation}
where $H^{jk}_{\ell,\sigma}(\{{ \B.r}_{m}\})$ and $H_{\ell,\sigma}(\{{
\B.r}_{m}\})$ are polynomials of degree $n$. The latter is fully
symmetric in the coordinates.  The former is symmetric in $r_j$, $r_k$
and separately in all the other $\{r_m\}_{m\ne i,j}$.

Substituting Eq.~(\ref{ansatzG}) into Eq.~(\ref{firsto}) and
collecting terms of the same type yields three equations:
\begin{eqnarray}
&&\sum_i \nabla_i^2 H^{jk}_{\ell,\sigma}=\nabla_j\cdot \nabla_k
E_{2n,\ell,\sigma}\ ,
\label{zero1}\\
&&2 \big[d-2+\B.r_{jk}\cdot
(\B.\nabla_j-\B.\nabla_k)\big]H^{jk}_{\ell,\sigma}\label{zero2}\\ 
&& \qquad +\frac{r_{jk}^\alpha r_{jk}^\beta \nabla_j^\alpha \nabla_k
^\beta}{d-1} E_{2n,\ell,\sigma} 
=-r_{jk}^2
K_{\ell,\sigma}^{jk}\ , \nn \\
&& \sum_i \nabla_i^2
H_{\ell,\sigma} =\sum_{j\ne k}K_{\ell,\sigma}^{jk} \ .
\label{zero3} 
\end{eqnarray}
Here $K_{\ell,\sigma}^{jk}$ are polynomials of degree $n-2$ which are
separately symmetric in the $j,k$ coordinates and in all the other
coordinates except $j,k$. In Ref.\cite{96BGK} it was proven that for
$\ell=0$ these equations possess a unique solution. The proof follows
through unchanged for any $\ell\ne 0$, and we thus proceed to finding
the solution.

By symmetry we can specialize the discussion to $j=1$, $k=2$. In light
of Eq.~(\ref{zero2}) we see that $H^{12}_{\ell,\sigma}$ must have at
least a quadratic contribution in $r_{12}$. This guarantees that
(\ref{ansatzG}) is nonsingular in the limit $r_{12}\to 0$.  The only
part of $H^{12}_{\ell,\sigma}$ that will contribute to structure
functions must contain $\B.r_3\dots \B.r_{n}$ at least once. Since
$H^{12}_{\ell,\sigma}$ has to be a polynomial of degree $n$ in the
coordinates, it must be of the form
\begin{equation}
H^{12}_{\ell,\sigma}=r_{12}^{\alpha_1}r_{12}^{\alpha_2}
r_3^{\alpha_3}\dots r_{n}^{\alpha_{n}} C^{\alpha_1\alpha_2\dots
\alpha_{n}} +[\dots]_{1,2} \ , \label{H}
\end{equation}
where $[\dots]_{1,2}$ contains terms with higher powers of $r_{12}$
and therefore do not contain some of the other coordinates $r_3\dots
r_{n}$. Obviously such terms are unimportant for the structure
functions. Since $H^{12}_{\ell,\sigma}$ has to be symmetric in
$\B.r_1,\B.r_2$ and $\B.r_3\dots \B.r_{n}$ separately, and it has to
belong to an $\ell,\sigma$ sector, we conclude that the constant
tensor $\B.C$ has must have the same symmetry and to belong to the
same sector. Consulting Appendix A, the most general form of $\B.C$ is
\BEA
C^{\alpha_1\alpha_2\dots \alpha_{2n}} &=&
aB_{2n,\ell,\sigma}^{\alpha_1\alpha_2\dots
\alpha_{2n}}+b\delta^{\alpha_1\alpha_2} B_{2n-2,\ell,\sigma}
^{\alpha_3\alpha_4\dots
\alpha_{2n}}\\ \nn 
&&+c\sum_{i\ne j>2}\delta
^{\alpha_1\alpha_i}\delta^{\alpha_2\alpha_j}
B_{2n-4,\ell,\sigma}^{\alpha_3\alpha_4\dots
\alpha_{2n}} \ .
\EEA
Substituting in Eq.~(\ref{zero2}) one finds
\BEA\label{interim}
(d+2)H^{12}_{\ell,\sigma}+\frac{r_{12}^{\alpha_1}
r_{12}^{\alpha_2}r_3^{\alpha_3}
\dots r_{2n}^{\alpha_{2n}}}{2d-2} B_{2n,\ell,\sigma}
^{\alpha_1\dots \alpha_{2n}}\\ \nn
+\case{1}{2}r_{12}^{\alpha_1}r_{12}^{\alpha_2}
\delta^{\alpha_1\alpha_2}K^{1,2}
_{\ell,\sigma}=[\dots]_{1,2} \ . 
\EEA
Substituting Eq.~(\ref{H}) and demanding that coefficients of the term
$r_1^{\alpha_1}\dots r_{n}^{\alpha_{n}}$ will sum up to zero, we
obtain
\BEA\label{aandc}
&&-2(d+2)a -\frac{2}{2d-2}=0 \ , \\ \nn
&& -2(d+2)c=0\ ; \qquad 
\Longrightarrow c=0 \ . 
\EEA
The coefficient $b$ is not determined from this equation due to
possible contributions from the unknown last term. We determine the
coefficient $b$ from Eq.~(\ref{zero1}).  After substituting the forms
we find
\BEA\nn 
&&4\delta^{\alpha_1\alpha_2} r_3^{\alpha_3}
\dots r_{2n}^{\alpha_{2n}}
[a B_{2n,\ell,\sigma}^{\alpha_1\dots \alpha_{2n}}+b\delta
^{\alpha_1\alpha_2} 
B_{2n-2,\ell,\sigma}^{\alpha_3\alpha_4
\dots\alpha_{2n}}]\\ 
&=&\delta^{\alpha_1\alpha_2}
r^{\alpha_3}\dots r^{\alpha_{2n}} B_{2n,\ell,\sigma}
^{\alpha_1\dots \alpha_{2n}}+
[\dots]_{1,2} \ .
\EEA
Recalling the identity (\ref{iden1}) we obtain
\begin{equation}
b=\frac{z_{n,\ell}}{4d}[1-4a] \ . \label{solb}
\end{equation}
Finally we find that $a$ is $n,\ell$-independent,
\begin{equation}
a=-\frac{1}{2(d+2)(d-1)}\ , \label{a}
\end{equation}
whereas $b$ does depend on $n$ and $\ell$, and we therefore
denote it as $b_{n,\ell}$
\begin{equation}
b_{n,\ell}=\frac{(d+1)}{4(d+2)(d-1)} z_{n,\ell} \ . \label{b}
\end{equation}
In the next Subsection  we compute from these results the scaling
exponents in  the sectors of the SO($d$) symmetry group with $\ell \le n$.
\subsection{The scaling exponents of the structure functions}
We now wish to show that the solution for the zero modes of the
correlation functions ${\cal F}_{n}$ (i.e $Z_{n}$) result in
homogeneous structure functions ${\cal S}_{n}$. In every sector
$\ell\le n,\sigma$ we compute the scaling exponents, and show that they are
independent of $\sigma$. Accordingly the scaling exponents are denoted
$\zeta_{n}^{\ell}$, and we compute them to first order in $\epsilon$.

Using (\ref{diffop},\,\ref{SinF}), the structure function is given by:
\end{multicols}
\widetext
\leftline{-------------------------------------------------------------------------}
\begin{eqnarray} 
{\cal S}_{n,\ell,\sigma}(\B.r_1,\Obr_1; \ldots ;
\B.r_{n},\Obr_{n} )&=& \Delta_1^{\alpha_1} \ldots
\Delta_{n}^{\alpha_{n}} \Btensor 
\nonumber\\ &&+
\epsilon \sum_{i \neq j}
\stackrel{\text{no } i,j} {\overbrace{ \Delta_1^{\alpha_1} \ldots
\Delta_{n}^{\alpha_{n}} } } f^{\alpha_i\alpha_j} (\B.r_i,\Obr_i,
\B.r_j,\Obr_j) [a\Btensor +b_{n,\ell}\Ctensor{i}{j}] \ ,
\label{Snunfused}
\end{eqnarray}
where $\Delta_i^{\alpha_i} \equiv r_i^{\alpha_i}-\Or_i^{\alpha_i}$, 
and the function $f$ is defined as:
\begin{eqnarray}
f^{\alpha_i\alpha_j} (\B.r_i,\Obr_i,\B.r_j,\Obr_j) & \equiv &
(r_i-r_j)^{\alpha_i}(r_i-r_j)^{\alpha_j} \ln|\B.r_i -\B.r_j| + (\Or_i
- \Or_j)^{\alpha_i}(\Or_i - \Or_j)^{\alpha_j} \ln|\Obr_i - \Obr_j| 
\\ \nn
&& - (r_i - \Or_j)^{\alpha_i}(r_i - \Or_j)^{\alpha_j} \ln |\B.r_i -
\Obr_j| - (\Or_i - r_j)^{\alpha_i}(\Or_i - r_j)^{\alpha_j}\ln
|\Obr_i-{\B.r}_j| \ .
\end{eqnarray} 
The scaling exponent of ${\cal S}_{n,\ell,\sigma}$ can be found by
multiplying all its coordinates by $\mu$. A direct calculation yields:
\begin{eqnarray*} 
&&{\cal S}_{n,\ell,\sigma}(\mu \B.r_1, \mu \Obr_1 ; \ldots ) =
\mu^{n}{\cal S}_{n,\ell,\sigma}(\B.r_1,\Obr_1 ; \ldots) \\
&-& 2\epsilon
\mu^{n}\ln \mu \sum_{i\neq j}\stackrel{\text{no } i,j} {
\overbrace{\Delta_1^{\alpha_1} \ldots \Delta_{n}^{\alpha_{n}} } }
\Delta_i^{\alpha_i}\Delta_j^{\alpha_j} [a\Btensor +
b_{n,\ell}\Ctensor{i}{j}] + O(\epsilon^2), \\ 
&=& \mu^{n}{\cal
S}_{n,\ell,\sigma}(\B.r_1,\Obr_1 ; \ldots ) - 2\epsilon \mu^{n} \ln
\mu \Delta_1^{\alpha_1} \ldots \Delta_{n}^{\alpha_{n}} 
\sum_{i\neq j} [a\Btensor + b_{n,\ell}\Ctensor{i}{j}] + O(\epsilon^2) \ .
\end{eqnarray*}
Using (\ref{iden2}), we find that
$$ \sum\limits_{i\neq j} [a\Btensor + b_{n,\ell}
\Ctensor{i}{j}] = [n(n-1)a+b_{n,\ell}]
\Btensor\,, 
$$
and therefore, we finally obtain:
\begin{eqnarray*}
{\cal S}_{n}(\mu \B.r_1, \mu \Obr_1 ;\ldots ) &=&
\mu^{n} \left\{ 1 - 2\epsilon [n(n-1)a+b_{n,\ell}] \ln \mu \right\} 
{\cal S}_{n}(\B.r_1, \Obr_1 ;\ldots ) + O(\epsilon^2) \\ &=&
\mu^{\zeta_{n}^{(\ell)}} {\cal S}_{n}(\B.r_1, \Obr_1 ; \ldots ) 
+ O(\epsilon^2)\ ,
\end{eqnarray*}
The result of the scaling exponent is now evident:
\begin{eqnarray}
\zeta_{n}^{(\ell)} &=&
n-2\epsilon[-\frac{n(n-1)}{2(d+2)(d-1)}+
\frac{(d+1)}{4(d+2)(d-1)}
z_{n,\ell}] +O(\epsilon^2) \nonumber\\
&=&n-\epsilon\Big[\frac{n(n+d)}{2(d+2)}
-\frac{(d+1)\ell(\ell+d-2)}{2(d+2)(d-1)}\Big]
+O(\epsilon^2) \ . \label{zetafinal}
\end{eqnarray} 
\rightline{-------------------------------------------------------------------------}
\begin{multicols}{2}
For $\ell=0$ this result coincides with \cite{96BGK}. This is the
final result of this calculation.  It is noteworthy that this result
is in full agreement with (\ref{Hfinal}) and (\ref{exp4}), even though
the scaling exponents that appear in these result refer to different
quantities. The way to understand this is the fusion rules that are
discussed next.
\subsection{Fusion Rules}
The fusion rules address the asymptotic properties of the fully
unfused structure functions when two or more of the coordinates are
approaching each other, whereas the rest of the coordinates remain
separated by much larger scales. A full discussion of the fusion rules
for the Navier-Stokes and the Kraichnan model can be found in
\cite{96FGLP,96LP}. In this section we wish to derive the fusion rules
directly from the zero modes that were computed to $O(\epsilon)$, in
all the sectors of the symmetry group. In other words, we are after
the dependence of the structure function $ {\cal S}_{n}({\bf
r}_{1},\overline{{\bf r}}_{1};\ldots )$ on its first $p$ pairs of
coordinates ${\bf r}_{1},\overline{{\bf r}}_{1};\ldots ;{\bf r}_{p},
\overline{{\bf r}}_{p}$ in the case where these points are very close
to each other compared to their distance from the other $n-p$ pairs of
coordinates. Explicitly, we consider the case where ${\bf r}_{1},
\overline{{\bf r}}_{1};\ldots ;{\bf r}_{p},\overline{{\bf r}}_{p}\ll
{\bf r} _{p+1},\overline{{\bf r}}_{p+1};\ldots ;{\bf
r}_{n},\overline{{\bf r}}_{n}$.  (We have used here the property of
translational invariance to put the center of mass of the first $2p$
coordinates at the origin). The calculation is presented in Appendix D,
with the final result (to $O(\epsilon)$)
\BEA\label{fusion}
&&{\cal S}_{n,\ell,\sigma}({\bf r}_{1},\overline{{\bf r}}_{1};\ldots
;{\bf r}_{n}, \overline{{\bf r}}_{n}) \\ \nn &=&\sum_{j=j_{\rm
max}}^p\sum_{\sigma ^{\prime }}\psi_{j,\sigma ^{\prime }}S_{p,j,
\sigma
^{\prime }}({\bf r}_{1},\overline{{\bf r}}_{1};\ldots ;{\bf
r}_{p},\overline{ {\bf r}}_{p}) \ .
\EEA
In this expression the quantity $\B.\psi_{j,\sigma ^{\prime }}$ is a
function of all the coordinates that remain separated by large
distances, and
\begin{equation}
j_{\rm max}={\rm max}\{0,p+\ell-n\} \ , \quad \ell \le n
\end{equation} 
We have shown that the LHS has a homogeneity exponent
$\zeta_{n}^{(\ell)}$. The RHS is a product of functions with
homogeneity exponents $\zeta_{p}^{(j)}$ and the functions
$\B.\psi_{j,\sigma ^{\prime}}$. Using the linear independence of the
functions $\B.S_{p,j,\sigma'}$ we conclude that $\B.\psi_{j,\sigma
^{\prime}}$ must have homogeneity exponent
$\zeta_{n}^{(\ell)}-\zeta_{p}^{(j)}$. This is precisely the prediction of
the fusion rules, but in each sector separately. One should stress the
intuitive meaning of the fusion rules. The result shows that when $p$
coordinates approach each other, the homogeneity exponent
corresponding to these coordinates becomes simply $\zeta_p^{(j)}$
as if we were considering a $p$-order correlation function. The
meaning of this result is that $p$ field amplitudes measured at $p$
close-by coordinates in the presence of $n-p$ field amplitudes
determined far away behave scaling-wise like $p$ field amplitudes in
the presence of anisotropic boundary conditions.  closing we note that
the tensor functions $\B.\psi_{j,\sigma ^{\prime}}$ do not necessarily
belong to the $j,\sigma'$ sector of SO($d$). 

\section{Summary and discussion}
One of our motivations in this paper was to understand the scaling
properties of the statistical objects under anisotropic boundary
conditions.  The scaling exponents were found for all $\ell\le n$,
cf. Eq.(1.1). We found a discrete and strictly increasing spectrum of
exponents as a function of $\ell$. This means that for higher $\ell$
the anisotropic contributions to the statistical objects decay faster
upon decreasing scales. In other words, the statistical objects tend
towards locally isotropic statistics upon decreasing the scale.  The
rate of isotropization is determined by the difference between the
$\ell$ dependent scaling exponents, and is of course a power law.  The
result shows that the $\ell$-dependent part is $n$-independent.  This
means that the rate of isotropization of all the moments of the
distribution function of field differences across a given scale is the
same. This is a demonstration of the fact that the distributions
function itself tends towards a locally isotropic distribution
function at the same rate. We note in passing that to first order in
$\epsilon$ the $\ell$ dependent part is also identical for $\zeta_2$,
a quantity whose isotropic value is {\em not} anomalous. For all
$\ell>1$ also $\zeta_2^{(\ell)}$ is anomalous, and in agreement with
the $n=1$ value of Eq.(1.1). Significantly, for $\zeta_2$ we have a
nonperturbative result that was derived in \cite{96FGLP}, namely
\BEA\label{zeta2}
\zeta_2^{(\ell)}&=&
\frac{1}{2}\Big[2-d-\epsilon\\ \nn 
&&+\sqrt{(2-d-\epsilon)^2
+\frac{4(d+\epsilon-1)\ell(d+\ell-2)}{d-1}}\Big] \ , 
\EEA
valid for all values of $\epsilon$ in the interval (0,2) and for all
$\ell\ge 2$.  This exact result agrees after expanding to
$O(\epsilon)$ with (\ref{zetafinal}) for $n=1$ and $\ell=2$ .

Our second motivation was to expose the correspondence between the
scaling exponents of the zero modes in the inertial interval and the
corresponding scaling exponents of the gradient fields. The latter do
not depend on any inertial scales, and the exponent appears in the
combination $\left(\Lambda/\xi\right)^{\zeta_{n}^\ell}$ where $\xi$ is
the appropriate ultraviolet inner cutoff, either $\lambda$ or $\eta$,
depending on the limiting process. We found exact agreement with the
exponents of the zero modes in all the sectors of the symmetry group
and for all values of $n$.  The deep reason behind this agreement is
the linearity of the fundamental equation of the passive scalar
(\ref{advect}). This translates to the fact that the viscous cutoff
$\eta$, Eq.(\ref{eta}) is $n$ and $\ell$ independent, and also does
not depend on the inertial separations in the unfused correlation
functions. This point has been discussed in detail in
\cite{96FGLP,96LPa}. In the case of Navier-Stokes statistics we expect
this ``trivial" correspondence to fail, but nevertheless the ``bridge
relations" that connect these two families of exponents has been
presented in \cite{96LP} for the isotropic sector. Finally we note
that in the present case we have displayed the fusion rules in all the
$\ell$ sectors, using the $O(\epsilon)$ explicit form of the zero
modes. We expect the fusion rules to have a nonperturbative validity
for any value of $\epsilon$. It would be interesting to explore
similar results for the Navier-Stokes case.

\acknowledgments
We want to thank  Yoram Cohen for numerous useful discussions.
This work has been supported in part by the Israel Science Foundation
administered by the Israel Academy of Sciences and Humanities, the
German-Israeli Foundation, The European Commission under the TMR
program, the Henri Gutwirth Fund for Research and the Naftali and Anna
Backenroth-Bronicki Fund for Research in Chaos and Complexity.

\appendix
\section{Anisotropy in d-dimensions}
To deal with anisotropy in d-dimensions we need classify the
irreducible representations of the group of all $d$-dimensional
rotations, SO($d$) \cite{62Ham}, and then to find a proper basis for
these representations.  The main linear space that we work in (the
carrier space) is the space of constant tensors with $n$ indices. This
space possesses a natural representation of SO($d$), given by the well
known transformation of tensors under $d$-dimensional rotation.

The traditional method to find a basis for the irreducible
representations of SO($d$) in this space, is using the Young tableaux
machinery on the subspace of traceless tensors \cite{62Ham}. It turns
out that in the context of the present paper, we do not need the
explicit structure of these tensors. Instead, all that matters are
some relations among them. A convenient way to derive these relations
is to construct the basis tensors from functions on the unit
$d$-dimensional sphere which belong to a specific irreducible
representation. Here also, the explicit form of these functions in
unimportant.  All that matters for our calculations is the action of
the Laplacian operator on these functions.

Let us therefore consider first the space ${\cal S}_d$ of functions
over the unit $d$-dimensional sphere. The representation of SO($d$)
over this space is naturally defined by:
\begin{equation}
    {\cal O}_{\cal R} \Psi(\hat{u}) \equiv \Psi({\cal R}^{-1}\hat{u}) \  ,
    \label{d-Rot}
\end{equation}
where $\Psi(\hat{u})$ is any function on the $d$-dimensional sphere, and
${\cal R}$ is a $d$-dimensional rotation.

${\cal S}_d$ can be spanned by polynomials of the unit vector
$\hat{u}$. Obviously (\ref{d-Rot}) does not change the degree of a
polynomial, and therefore each irreducible representation in this
space can be characterized by an integer $\ell=0,1,2,\ldots$,
specifying the degree of the polynomials that span this
representation. At this point, we cannot rule out the possibility that
some other integers are needed to fully specify all irreducible
representations in ${\cal S}_d$ and therefore we will need below
another set of indices to complete the specification.

We can now choose a basis of polynomials $\{ Y_{\ell,\sigma}(\hat u)
\}$ that span all the irreducible representations of SO($d$) over
${\cal S}_d$. The index $\sigma$ counts all integers other than $\ell$
needed to fully specify all irreducible representations, and in
addition, it labels the different functions within each irreducible
representation.

Let us demonstrate this construction in two and three dimensions. In
two dimensions $\sigma$ is unneeded since all the irreducible
representation are one-dimensional and are spanned by $Y_\ell(\hat u)=
e^{i\ell\phi}$ with $\phi$ being the angle between $\hat u$ and the
the vector $\hat e_1\equiv (1,0)$.  Any rotation of the coordinates in
an angle $\phi_0$ results in a multiplicative factor $e^{i\phi_0}$. It
is clear that $Y_\ell(\hat u)$ is a polynomial in $\hat u$ since
$Y_\ell(\hat u)=[\hat u\cdot \hat p]^\ell$ where $\hat p\equiv (1,i)$.
In three dimensions $\sigma=m$ where $m$ takes on $2\ell+1$ values
$m=-\ell, -\ell+1,\dots,\ell$. Here $Y_{\ell,m}\propto e^{im\phi}
P^m_\ell(\cos\theta)$ where $\phi$ and $\theta$ are the usual
spherical coordinates, and $P^m_\ell$ is the associated Legendre
polynomial of degree $\ell-m$. Obviously we again have a polynomial in
$\hat u$ of degree $\ell$.

We now wish to calculate the action of the Laplacian operator with
respect to $u$ on the $Y_{\ell,\sigma}(\hat u)$. We prove the
following identity:
\begin{equation}
u^2\partial^\alpha\partial_\alpha Y_{\ell,\sigma}(\hat u) =
-\ell(\ell+d-2)Y_{\ell,\sigma}(\hat u) \ . \label{LapY}
\end{equation}
One can easily check that for $d=3$ (\ref{LapY}) gives the factor
$\ell (\ell + 1)$, well known from the theory of angular-momentum in
Quantum Mechanics. To prove this identity for any $d$, note that
\begin{equation}
|u|^{2-\ell}\partial^2 |u|^\ell Y_{\ell,\sigma}(\hat u)) =0 \ .
\label{firststep}
\end{equation}
This follows from the fact that the Laplacian is an isotropic
operator, and therefore is diagonal in the $Y_{\ell,\sigma}$. The same
is true for the operator $|u|^{2-\ell}\partial^2 |u|^\ell$.  But this
operator results in a polynomial in $\hat u$ of degree $\ell-2$, which
is spanned by $Y_{\ell',\sigma'}$ such that
$\ell'\le\ell-2$. Therefore the RHS of (\ref{firststep}) must
vanish. Accordingly we write
\begin{equation}
\partial^2|u^\ell|Y_{\ell,\sigma}(\hat u)+2\partial^\alpha|u^\ell|\partial^\alpha
Y_{\ell,\sigma}+|u^\ell|\partial^2 Y_{\ell,\sigma}(\hat u) = 0 \ .
\end{equation}
The second term vanishes since it contains a radial derivative
$u^\alpha\partial_\alpha$ operating on $Y_{\ell,\sigma}(\hat u)$ which
depends on $\hat u$ only. The first and third terms, upon elementary
manipulations, lead to (\ref{LapY}).

Having the $Y_{\ell,\sigma}(\hat u)$ we can now construct the
irreducible representations in the space of constant tensors. The
method is based on acting on the $Y_{\ell,\sigma}(\hat u)$ with the
{\em isotropic} operators $u^\alpha,~\partial^\alpha$ and
$\delta^{\alpha\beta}$. Due to the isotropy of the above operators,
the behavior of the resulting expressions under rotations is similar
to the behavior of the scalar function we started with. For example,
the tensor fields $\delta^{\alpha\beta}Y_{\ell,\sigma}(\hat u),
\partial^{\alpha}\partial^{\beta}Y_{\ell,\sigma}(\hat u) $ transform
under rotations according to the $(\ell, \sigma)$ sector of SO($d$).

Next, we wish to find the basis for the irreducible representations of the
space of constant and fully symmetric tensors with $n$ indices. We form
the basis
\begin{equation}
B^{\alpha_1,\dots,\alpha_{n}}_{\ell,\sigma,n}\equiv \partial^{\alpha_1}
\dots \partial^{\alpha_{n}} u^{n} Y_{\ell,\sigma}(\hat u) , \quad
\ell\le n \ .
\label{Birr}
\end{equation}
Note that when $\ell$ {\em and} $n$ are even (as is the case
invariably in this paper),
$B^{\alpha_1,\dots,\alpha_{n}}_{\ell,\sigma,n}$ no longer depends on
$\hat u$, and is indeed fully symmetric by construction. Simple
arguments can also prove that this basis is indeed complete, and spans
{\em all} fully symmetric tensors with $n$ indices. Other examples of
this procedure for the other spaces are presented directly in the
text.

Finally let us introduce two identities involving the
$B_{n,\ell,\sigma}$ which are used over and over through the
paper. The first one is
\begin{eqnarray}
\delta_{\alpha_1\alpha_2}B^{\alpha_1,\dots,\alpha_{n}}_{\ell,\sigma,n}
&=&z_{n,\ell}B^{\alpha_3,\dots,\alpha_{n}}_{\ell,\sigma,n-2} \ ,
\label{iden1}\\ z_{n,\ell}&=&[n(n+d-2)-\ell(\ell+d-2)] \ . \label{znl}
\end{eqnarray}
It is straightforward to derive this identity using (\ref{LapY}). The
second identity is
\begin{equation}
\sum_{i\ne j} \delta^{\alpha_i\alpha_j}B^{\{\alpha_m\},m\ne
i,j}_{\ell,\sigma,n-2}
=B^{\alpha_1,\dots,\alpha_{n}}_{\ell,\sigma,n} \ , \quad \ell\le n-2 \ .
\label{iden2}
\end{equation}
This identity is proven by writing $u^{n}$ in (\ref{Birr}) as $u^2
u^{n-2}$, and operating with the derivative on $u^2$. The term
obtained as $u^2\partial^{\alpha_1} \dots \partial^{\alpha_{n}}
u^{n-2} Y_{\ell,\sigma}(\hat u) $ vanishes because we have $n$
derivatives on a polynomial of degree $n-2$. It is worthwhile noticing
that these identities connect tensors from two different spaces. The
space of tensors with $n$ indices and the space of tensors with $n-2$
indices.  Nevertheless, in both spaces, the tensors belong to the same
$(\ell, \sigma)$ sector of the SO($d$) group. This is due to the
isotropy of the contraction with $\delta^{\alpha_1\alpha_2}$ in the
first identity, and the contraction with $\delta^{\alpha_i\alpha_j}$
in the second identity.

\section{Proof of Eq.~(3.32)}

In case a of region II where $p_1>p_2>k_s$ the analytic expression for
$\B.A_2$ can be simplified to
\begin{equation}
A_{2.\beta_i\beta_j}^{\alpha_i\alpha_j}=\frac{1}{2}\hat p_1^{\alpha_i}
\hat p_1^{\alpha_j}\int \frac{d\hat
p_2}{\Omega_d}P^{\beta_i\beta_j}(\hat p_2) \ .
\end{equation}
Using the identities
\begin{eqnarray}
\int d\hat p &=& \Omega(d) \ ,\quad \Omega_d\equiv
\Omega(d)\frac{d-1}{d}\ , \\ \int d\hat p \hat p^\alpha \hat p^\beta
&=&\delta_{\alpha\beta}\Omega(d) \ ,
\end{eqnarray}
we compute
\begin{equation}
A_{2.\beta_i\beta_j}^{\alpha_i\alpha_j}=\case{1}{2}\hat p_1^{\alpha_i}
\hat p_1^{\alpha_j}\delta_{\beta_i\beta_j} \ .
\end{equation}
Substituting this form into the double rung ladder diagram results,
after contracting all the indices of $\B.A_1$ and $\B.A_2$, in a form
identical to Eq.~(\ref{int1}) for $\B.A$ in the one rung ladder
diagram. This leads directly to the final equation Eq.~(\ref{casea}).
\section{Double resummation}
\subsection{Calculation of $D_{m,s}$}
In this Appendix we discuss the calculation of the coefficients
$D_{m,s}$ in Eq.~(\ref{Ksum}), and the actual resummation of that
equation.

Firstly we need to introduce rules to evaluate the rungs in the
general ladder diagram that appears in the expansion. The rule is
actually quite simple: every rung contributes a term proportional to
$gA_{n}^{(\ell)}$ if the $\B.p$ vector associated with this rung is
the largest among all the $\B.p$ vectors associated with rungs
appearing to the right of it.  Otherwise the contribution is
proportional to $\tilde g$. The weight of the contribution is obtained
as a factor $c\le 1$ which reflects the proportional fraction of the
volume of $(\B.p_1,\B.p_2 \dots)$-space in which the associated
ordering of the $\B.p$ vectors is valid. For example, if the rung with
the largest $\B.p$ vector is in the extreme right, then all the other
rungs contribute terms proportional to $\tilde g$. Thus a diagram with
$s$ rungs ordered in this manner contributes with a weight
$C=1/s$. Therefore
\begin{equation}
D_{1,s}=\frac{1}{s} \  .  \label{C1}
\end{equation}
The recurrence relation for $D_{m,s}$ with $m>1$ is derived by
inserting an additional rung which is associated with the largest
$\B.p$ vector in any one of the $(s+1)$ possible positions available
in a diagram with $s$ rungs. After some combinatorial calculations of
the weights one finds
\begin{equation}
D_{m+1,s}=\frac{1}{s}\sum_{q=m}^{s-1}D_{q,m} . 
\end{equation}
Together with (\ref{C1}) this gives

\begin{eqnarray}
D_{2,s}&=&\frac{1}{s}\sum_{q=1}^{s-1}\frac{1}{q} \ , \\
D_{3,s}&=&\frac{1}{s}\sum_{q_2=2}^{s-1}\frac{1}{q_2}
\sum_{q_1=1}^{q_2-1}\frac{1}{q_1} \ ,\\ 
D_{4,s}&=&\frac{1}{s}\sum_{q_3=3}^{s-1}\frac{1}{q_3}
\sum_{q_2=2}^{q_3-1}\frac{1}{q_2}\sum_{q_1=1}
^{q_2-1}\frac{1}{q_1} \ , \quad etc. 
\end{eqnarray}
The general structure of $D_{m,s}$ now becomes obvious.
\subsection{Higher order terms in $A$}

Consider the equation (\ref{Kn2})
\begin{equation}
K_2(g,A)=A^2\sum_{s=2}^\infty \frac{\tilde g^s}{s}
\sum_{q=1}^{s-1}\frac{1}{q}
\ . \label{C2}
\end{equation}
Observing that 
\begin{equation}
\sum_{q=1}^{s-1} \frac{1}{q} 
= \frac{1}{2}\Big[ \sum_{q=1}^{s-1}\left(\frac{1}{s-q}
+\frac{1}{q}\right)\Big] \ , \label{trick1}
\end{equation}
and 
\begin{equation}
\frac{1}{s}\left(\frac{1}{q}
+\frac{1}{s-q}\right) 
= \frac{1}{q(s-q)} \ , \label{trick2}
\end{equation}
we end up with
\begin{equation}
K_2(\tilde g, A)=\frac{1}{2} A^2
\sum_{q_1=1}^\infty\sum_{q_2=1}^\infty 
\frac{\tilde g^{q_1+q_2}}{q_1q_2} \ ,
\end{equation}
where we relabeled $q\to q_1$ and $s-q\to q_2$ and changed
correspondingly the limit of summation over $q_2$. Thus
\begin{equation}
K_2(\tilde g,A)=\frac{1}{2} A^2\Big [\sum_{q=1}^\infty
\frac{1}{q}\Big]^2=
\frac{1}{2}\Big[-A\ln(1-g)\Big]^2 \ .
\end{equation}
The terms proportional to $A^3$ give
\begin{equation}
K_3(\tilde g,A)=A^3\sum_{s=3}^\infty 
\frac{\tilde g^n}{n}\sum_{q_2=2}^{s-1}
\frac{1}{q_2}\sum_{q_1=1}^{q_2-1}\frac{1}{q_1}
\ . \label{C4}
\end{equation}
We can rearrange the sums by summing over $q_3=n-q_1-q_2$ instead of
$n$.  Using relationships similar to (\ref{trick1}) and (\ref{trick2})
we find
\begin{equation}
K_3=\frac{1}{6} A^2\sum_{q_1=1}^\infty\sum_{q_2=1}
^\infty 
\sum_{q_3=1}^\infty\frac{\tilde g^{q_1+q_2+q_3}}{q_1q_2q_3} \ .
\end{equation}
Obviously this leads to 
\begin{equation}
K_3(\tilde g,A)=\frac{1}{6} 
A^3\Big [\sum_{q=1}^\infty \frac{\tilde g^q}{q}\Big]^3=
\frac{1}{3!}\Big[-A\ln(1-\tilde g)\Big]^3 \ .
\end{equation}
The general structure is now clear, leading to Eq.~(\ref{Kn2final}).
\section{Derivation of the fusion rules}
In this appendix we derive the fusion rules (\ref{fusion}). Consider a
fully unfused structure function with $n$ coordinates, such that $p$
of them are separated from each other by a typical distance $r$,
whereas $n-p$ coordinates are separated from them and from each other
by a typical distance $R$, and $R\gg r$. We want to compute the
asymptotic properties of $S_n$ and show that to leading order in $r/R$
we find Eq.(\ref{fusion}). For homogeneous ensembles we can shift the
origin to the center of mass of the $p$ coordinates. In this case we
have $r_j\ll r_i$ for every $j\le p$ and $i>p$.  Our aim is to
separate the dependence on the small distances from the dependence on
the large distances. We will see that some of the terms in $S_n$ lend
themselves naturally to such a separation, and some call for more
work. We start from Eq.(\ref{Snunfused}), and compute to first order
in $r/R$:
\end{multicols}
\leftline{--------------------------------------------------------------------------}
\begin{eqnarray}
f^{\alpha _{i}\alpha _{j}}({\bf r}_{i},\overline{{\bf r}}_{i},{\bf
r}_{j},
\overline{{\bf r}}_{j}) &\equiv &(r_{i}-r_{j})^{\alpha
_{i}}(r_{i}-r_{j})^{\alpha _{j}}\ln |{\bf r}_{i}-{\bf
r}_{j}|+(\overline{r} _{i}-\overline{r}_{j})^{\alpha
_{i}}(\overline{r}_{i}-\overline{r} _{j})^{\alpha _{j}}\ln
|\overline{{\bf r}}_{i}-\overline{{\bf r}}_{j}| \nonumber\\
&&-(r_{i}-\overline{r}_{j})^{\alpha
_{i}}(r_{i}-\overline{r}_{j})^{\alpha _{j}}\ln |{\bf
r}_{i}-\overline{{\bf r}}_{j}|-(\overline{r} _{i}-r_{j})^{\alpha
_{i}}(\overline{r}_{i}-r_{j})^{\alpha _{j}}\ln |
\overline{{\bf r}}_{i}-{\bf r}_{j}|\; \nonumber\\
&=&[-2r_{i}^{\alpha _{i}}\delta ^{\alpha _{j}}{}_{\beta }\ln
r_{i}-r_{i}^{\alpha _{i}}r_{i}^{\alpha _{i}}\frac{(r_{i})_{\beta
}}{r_{i}^{2} }+2\overline{r}_{i}^{\alpha _{i}}\delta ^{\alpha
_{j}}{}_{\beta }\ln
\overline{r}_{i}+\overline{r}_{i}^{\alpha _{i}}\overline{r}_{i}^{\alpha _{i}}
\frac{(\overline{r}_{i})_{\beta }}{\overline{r}_{i}^{2}}]\Delta _{j}^{\beta }
\nonumber\\
&\equiv &g^{\alpha _{i}\alpha _{j}}{}_{\beta }({\bf
r}_{i},\overline{{\bf r}} _{i})\Delta _{j}^{\beta } \ ,
\label{fexpand}
\end{eqnarray}
and so, if $r_{j},\overline{r}_{j}\ll r_{i},\overline{r}_{i}$ for $
j=1,\ldots ,p\;,\;i=p+1,\ldots ,n$ then the first order in $\epsilon$ of ${\cal S}_{n}$ will
contain 3 types of terms:
\begin{eqnarray*}
I_{1} &=&\sum_{1\leq i\neq j\leq p}\stackrel{\text{no }i,j}{\overbrace{
\Delta _{1}^{\alpha _{1}}\ldots \Delta _{n}^{\alpha _{n}}}}f^{\alpha
_{i}\alpha _{j}}({\bf r}_{i},\overline{{\bf r}}_{i},{\bf r}_{j},\overline{
{\bf r}}_{j})[aB_{n,\ell ,\sigma }^{\alpha _{1}\ldots \alpha _{n}}+b_{n,\ell}\delta
^{\alpha _{i}\alpha _{j}}\stackrel{\text{no }i,j}{\overbrace{B_{n-2,\ell
,\sigma }^{\alpha _{1}\ldots \alpha _{n}}}]} \\
&=&\sum_{1\leq i\neq j\leq p}\stackrel{\text{no }i,j}{\overbrace{\Delta
_{1}^{\alpha _{1}}\ldots \Delta _{p}^{\alpha _{p}}}}f^{\alpha _{i}\alpha
_{j}}({\bf r}_{i},\overline{{\bf r}}_{i},{\bf r}_{j},\overline{{\bf r}}
_{j})[aB_{n,\ell ,\sigma }^{\alpha _{1}\ldots \alpha _{n}}+b_{n,\ell}\delta
^{\alpha _{i}\alpha _{j}}\stackrel{\text{no }i,j}{\overbrace{B_{n-2,\ell
,\sigma }^{\alpha _{1}\ldots \alpha _{n}}}]}\Delta _{p+1}^{\alpha
_{p+1}}\ldots \Delta _{n}^{\alpha _{n}}
\end{eqnarray*}
\begin{eqnarray*}
I_{2} &=&\sum_{1\leq j\leq p,p<i\leq n}\stackrel{\text{no }i,j}{\overbrace{
\Delta _{1}^{\alpha _{1}}\ldots \Delta _{n}^{\alpha _{n}}}}f^{\alpha
_{i}\alpha _{j}}({\bf r}_{i},\overline{{\bf r}}_{i},{\bf r}_{j},\overline{
{\bf r}}_{j})[aB_{n,\ell ,\sigma }^{\alpha _{1}\ldots \alpha _{n}}+b_{n,\ell}\delta
^{\alpha _{i}\alpha _{j}}\stackrel{\text{no }i,j}{\overbrace{B_{n-2,\ell
,\sigma }^{\alpha _{1}\ldots \alpha _{n}}}]} \\
&=&\Delta _{1}^{\alpha _{1}}\ldots \Delta _{p}^{\alpha _{p}}\sum_{1\leq
j\leq p,p<i\leq n}\stackrel{\text{no }i}{\overbrace{\Delta _{p+1}^{\alpha
_{p+1}}\ldots \Delta _{n}^{\alpha _{n}}}}g^{\alpha _{i}\beta }{}_{\alpha
_{j}}({\bf r}_{i},\overline{{\bf r}}_{i})[aB_{n,\ell ,\sigma }^{\alpha
_{1}\ldots \beta \ldots \alpha _{n}}+b_{n,\ell}\delta ^{\alpha _{i}\beta }\stackrel{
\text{no }i,j}{\overbrace{B_{n-2,\ell ,\sigma }^{\alpha _{1}\ldots \alpha
_{n}}}]}
\end{eqnarray*}

\begin{eqnarray*}
I_{3} &=&\sum_{p<i,j\leq n}\stackrel{\text{no }i,j}{\overbrace{\Delta
_{1}^{\alpha _{1}}\ldots \Delta _{n}^{\alpha _{n}}}}f^{\alpha _{i}\alpha
_{j}}({\bf r}_{i},\overline{{\bf r}}_{i},{\bf r}_{j},\overline{{\bf r}}
_{j})[aB_{n,\ell ,\sigma }^{\alpha _{1}\ldots \alpha _{n}}+b_{n,\ell}\delta
^{\alpha _{i}\alpha _{j}}\stackrel{\text{no }i,j}{\overbrace{B_{n-2,\ell
,\sigma }^{\alpha _{1}\ldots \alpha _{n}}}]} \\
&=&\Delta _{1}^{\alpha _{1}}\ldots \Delta _{p}^{\alpha _{p}}\sum_{p<i,j\leq
n}\stackrel{\text{no }i,j}{\overbrace{\Delta _{p+1}^{\alpha _{p+1}}\ldots
\Delta _{n}^{\alpha _{n}}}}f^{\alpha _{i}\alpha _{j}}({\bf r}_{i},
\overline{{\bf r}}_{i},{\bf r}_{j},\overline{{\bf r}}_{j})[aB_{n,\ell
,\sigma }^{\alpha _{1}\ldots \alpha _{n}}+b_{n,\ell}\delta ^{\alpha _{i}\alpha _{j}}
\stackrel{\text{no }i,j}{\overbrace{B_{n-2,\ell ,\sigma }^{\alpha
_{1}\ldots \alpha _{n}}}]} \ .
\end{eqnarray*}
We note that of these three terms only $I_2$ has a nontrivial mixing of
small and large coordinates, and indeed it is the only term in which
the expansion (\ref{fexpand}) was employed. Collecting terms we find 
\begin{eqnarray*}
&& {\cal S}_{n}({\bf r}_{1},\overline{{\bf r}}_{1};\ldots ;{\bf r}_{n},
\overline{{\bf r}}_{n}) =\Delta _{1}^{\alpha _{1}}\ldots
\Delta _{p}^{\alpha _{p}}\tilde{B}_{p}^{\alpha _{1}\ldots \alpha _{p}}
\\
&&+\epsilon
\sum_{i\neq j}\stackrel{\text{no }i,j}{\overbrace{\Delta _{1}^{\alpha
_{1}}\ldots \Delta _{p}^{\alpha _{p}}}}f^{\alpha _{i}\alpha _{j}}({\bf
r} _{i},\overline{{\bf r}}_{i},{\bf r}_{j},\overline{{\bf r}}_{j})
\lbrack a\tilde{B}_{p}^{\alpha _{1}\ldots \alpha _{p}}+b_{n,l}\delta
^{\alpha _{i}\alpha _{j}}\stackrel{\text{no }i,j}{
\overbrace{\tilde{B}_{p-2}^{\alpha _{1}\ldots \alpha _{p}}}]}
+\epsilon \Delta _{1}^{\alpha _{1}}\ldots \Delta _{p}^{\alpha _{p}}
\tilde{C}_{p}^{\alpha _{1}\ldots \alpha _{p}}+O(\epsilon ^{2}) ,
\end{eqnarray*}
where: 
\begin{eqnarray*}
\tilde{B}_{p}^{\alpha _{1}\ldots \alpha _{p}} &=&\Delta _{p+1}^{\alpha
_{p+1}}\ldots \Delta _{n}^{\alpha _{n}}B_{n,\ell ,\sigma }^{\alpha
_{1}\ldots \alpha _{n}}=\sum_{j={\rm max}\{0,p+\ell
-n\}}^{p}\sum_{\sigma ^{\prime }}c_{j,\sigma ^{\prime }}B_{p,j,\sigma
^{\prime }}^{\alpha _{1}\ldots \alpha _{p}} \\ \tilde{B}_{p-2}^{\alpha
_{3}\ldots \alpha _{p}} &=&\Delta _{p+1}^{\alpha _{p+1}}\ldots \Delta
_{n}^{\alpha _{n}}B_{n-2,\ell ,\sigma }^{\alpha _{3}\ldots \alpha
_{n}}=\sum_{j={\rm max}\{0,p+\ell -n\}}^{p-2}
\sum_{\sigma ^{\prime
}}d_{j,\sigma ^{\prime }}B_{p-2,j,\sigma ^{\prime }}^{\alpha
_{1}\ldots \alpha _{p}} \\ \tilde{C}_{p}^{\alpha _{1}\ldots \alpha
_{p}} &=&\sum_{p<i,j\leq n} \stackrel{\text{no }i,j}{\overbrace{\Delta
_{p+1}^{\alpha _{p+1}}\ldots \Delta _{n}^{\alpha _{n}}}}f^{\alpha
_{i}\alpha _{j}}({\bf r}_{i}, \overline{{\bf r}}_{i},{\bf
r}_{j},\overline{{\bf r}}_{j})[aB_{n,\ell ,\sigma }^{\alpha _{1}\ldots
\alpha _{n}}+b_{n,\ell}\delta ^{\alpha _{i}\alpha _{j}}
\stackrel{\text{no }i,j}{\overbrace{B_{n-2,\ell ,\sigma }^{\alpha
_{1}\ldots \alpha _{n}}}]} \\ &&+\sum_{1\leq j\leq p,p<i\leq
n}\stackrel{\text{no }i}{\overbrace{\Delta _{p+1}^{\alpha
_{p+1}}\ldots \Delta _{n}^{\alpha _{n}}}}g^{\alpha _{i}\beta
}{}_{\alpha _{j}}({\bf r}_{i},\overline{{\bf r}}_{i})[aB_{n,\ell
,\sigma }^{\alpha _{1}\ldots \beta \ldots \alpha
_{n}}+b_{n,\ell}\delta ^{\alpha _{i}\beta }\stackrel{\text{no
}i,j}{\overbrace{B_{n-2,\ell ,\sigma }^{\alpha _{1}\ldots \alpha
_{n}}}]} \\ &=&\sum_{j={\rm max}\{0,p+\ell -n\}}^{p}\sum_{\sigma
^{\prime }}e_{j,\sigma ^{\prime }}B_{p,j,\sigma ^{\prime }}^{\alpha
_{1}\ldots \alpha _{p}} \ .
\end{eqnarray*}
In these expressions we use the fact $\tilde\B.B_p$ is a fully
symmetric tensor and therefore can be again expanded in terms of the
basis functions $\B.B_{p,j,\sigma'}$ with coefficients that depend on
the large separations. The sums on the right hand sides run between
${j={\rm max}\{0,p+\ell -n\}}$ and $p$ because not all the basis
functions can appear when $p+\ell -n>0$. This can be checked by
contracting the $\tilde\B.B_p$ with $(n-\ell+2 )/2$ delta
function. This contraction vanishes since it contains a factor
$z_{\ell,\ell}$. On the other hand the contraction results in a tensor
with $p+\ell-n-2$ indices, and therefore all corresponding
coefficients of $j\le p+\ell-n-2$ must vanish. To proceed we establish
the following identity:
\begin{equation}
\frac{z_{p,j}}{z_{n,\ell}}c_{j,\sigma ^{\prime }}=d_{j,\sigma ^{\prime }}
\label{eq:key}
\end{equation}
The identity is proven by the following calculations: 
\begin{eqnarray*}
\Delta _{p+1}^{\alpha _{p+1}}\ldots \Delta _{n}^{\alpha _{n}}B_{n,\ell
,\sigma }^{\alpha _{1}\ldots \alpha _{n}} &=&\partial ^{\alpha
_{1}}\ldots \partial ^{\alpha _{p}}u^{p}u^{-p}[\Delta _{p+1}^{\alpha
_{p+1}}\ldots \Delta _{n}^{\alpha _{n}}]\partial _{\alpha
_{p+1}}\ldots \partial _{\alpha _{n}}u^{n}Y_{\ell ,\sigma }(\hat{u})
\\ &=&\partial ^{\alpha _{1}}\ldots \partial ^{\alpha
_{p}}u^{p}\sum_{j,\sigma ^{\prime }}c_{j^,\sigma ^{\prime
}}Y_{j,\sigma ^{\prime }}(\hat{u}) \end{eqnarray*} \begin{eqnarray*}
\Delta _{p+1}^{\alpha _{p+1}}\ldots \Delta _{n}^{\alpha
_{n}}B_{n-2,\ell ,\sigma }^{\alpha _{3}\ldots \alpha _{n}} &=&\partial
^{\alpha _{3}}\ldots \partial ^{\alpha _{p}}u^{p-2}u^{-p+2}[\Delta
_{p+1}^{\alpha _{p+1}}\ldots \Delta _{n}^{\alpha _{n}}]\partial
_{\alpha _{p+1}}\ldots \partial _{\alpha _{n}}u^{n-2}Y_{\ell ,\sigma
}(\hat{u}) \\ &=&\partial ^{\alpha _{3}}\ldots \partial ^{\alpha
_{p}}u^{p}\sum_{j,\sigma ^{\prime }}d_{j,\sigma ^{\prime }}Y_{j,\sigma
^{\prime }}(\hat{u}) \ .
\end{eqnarray*}
Denote now
\begin{eqnarray*}
f(\hat{u}) &\equiv &u^{-p}[\Delta _{p+1}^{\alpha _{p+1}}\ldots \Delta
_{n}^{\alpha _{n}}]\partial _{p+1}\ldots \partial _{n}u^{n}Y_{\ell
,\sigma }(\hat{u})=\sum_{j,\sigma ^{\prime }}c_{j,\sigma ^{\prime
}}Y_{j,\sigma ^{\prime }}(\hat{u}) \\ g(\hat{u}) &\equiv
&\;u^{-p+2}[\Delta _{p+1}^{\alpha _{p+1}}\ldots \Delta _{n}^{\alpha
_{n}}]\partial _{p+1}\ldots \partial _{n}u^{n-2}Y_{\ell ,\sigma
}(\hat{u})=\sum_{j,\sigma ^{\prime }}d_{j,\sigma ^{\prime
}}Y_{j,\sigma ^{\prime }}(\hat{u})
\end{eqnarray*}
To obtain (\ref{eq:key}), operate with $u^{2}\partial ^{2}$ on
$f(\hat{u})$.  On one hand, we get:
\[
u^{2}\partial ^{2}f(\hat{u})=\sum_{j,\sigma 
^{\prime }}-j(j+d-2)c_{j,\sigma ^{\prime
}}Y_{j,\sigma ^{\prime }}(\hat{u})
\]
but on the other hand, we have: 
\begin{eqnarray*}
u^{2}\partial ^{2}f(\hat{u}) &=&u^{2}\partial ^{2}[u^{-p}[\Delta
_{p+1}^{\alpha _{p+1}}\ldots \Delta _{n}^{\alpha _{n}}]\partial
_{p+1}\ldots \partial _{n}u^{n}Y_{\ell ,\sigma }(\hat{u})] \\
&=&-p(-p+d-2)f(\hat{u})-2p^{2}f(\hat{u})+z_{n,\ell}g(\hat{u}) \\
&=&-p(p+d-2)f(\hat{u})+z_{n,\ell}g(\hat{u}) \ .
\end{eqnarray*}
Equating the two expressions, and projecting over the $(j,\sigma
^{\prime })$ sector, we obtain: 
\begin{eqnarray*}
-j(j+d-2)c_{j,\sigma ^{\prime }}
&=&-p(p+d-2)c_{j,\sigma'}+z_{n,l}d_{j,\sigma ^{\prime }} \\
d_{j,\sigma ^{\prime }}
&=&\frac{[p(p+d-2)-j(j+2-2)]}{z_{n,l}}c_{j,\sigma ^{\prime
}}=\frac{z_{p,j}}{ z_{n,l}}c_{j,\sigma ^{\prime }}.
\end{eqnarray*}
Recalling Eq.(\ref{b}), $b_{p,j}=b_{n,l} z_{p,j}/{z_{n,l}}$ and
we may write to leading order in ${r}/{R}$:
\begin{eqnarray*}
&&{\cal S}_{n}({\bf r}_{1},\overline{{\bf r}}_{1};\ldots ;{\bf r}_{n},
\overline{{\bf r}}_{n}) =\sum^p_{j={\rm
max}\{0,p+\ell-n\}}\sum_{\sigma'}c_{j,\sigma'}\Big[\Delta _{1}^{\alpha
_{1}}\ldots \Delta _{p}^{\alpha _{p}}B_{p,j,\sigma'}^{\alpha
_{1}\ldots \alpha _{p}}\nonumber\\ &&+\epsilon \sum_{i\neq
j}\stackrel{\text{no }i,j}{\overbrace{\Delta _{1}^{\alpha _{1}}\ldots
\Delta _{p}^{\alpha _{p}}}}f^{\alpha _{i}\alpha _{j}}({\bf r}
_{i},\overline{{\bf r}}_{i},{\bf r}_{j},\overline{{\bf r}}_{j})
\lbrack a{B}_{p,j,\sigma'}^{\alpha _{1}\ldots \alpha
_{p}}+b_{p,j}\delta ^{\alpha _{i}\alpha _{j}}\stackrel{\text{no }i,j}{
\overbrace{{B}_{p-2,j,\sigma'}^{\alpha _{1}\ldots \alpha
_{p}}}]}\Big]\; \\ &&+ \sum^p_{j={\rm
max}\{0,p+\ell-n\}}\sum_{\sigma'}\epsilon e_{j,\sigma'}\Delta
_{1}^{\alpha _{1}}\ldots \Delta _{p}^{\alpha _{p}}
B_{p,j,\sigma'}^{\alpha _{1}\ldots \alpha _{p}}+O(\epsilon
^{2})\nonumber\\ &&=\sum^p_{j={\rm
max}\{0,p+\ell-n\}}\sum_{\sigma'}\left(c_{j,\sigma'}+\epsilon
e_{j,\sigma'}\right) S_{p,j,\sigma'}({\bf r}_{1},\overline{{\bf
r}}_{1};\ldots ;{\bf r}_{p}, \overline{{\bf r}}_{p})+O(\epsilon ^{2})
\ .
\end{eqnarray*}
From this follows Eq.(\ref{fusion}).
\begin{multicols}{2}

\end{multicols}
\end{document}